\documentclass[twocolumn,tighten]{aastex61}
\pdfoutput=1
\usepackage{amsmath,amstext}
\usepackage[T1]{fontenc}
\usepackage{apjfonts}
\usepackage[figure,figure*]{hypcap}

\newcommand*{\putfig}[1]{\centering{\includegraphics{figure_#1}}}

\newcommand*{\sub}[2]{{#1}_{\text{#2}}}
\newcommand*{\wprp}{w_p(r_p)}
\newcommand*{\Gpch}{\,\text{Gpc}\,h^{-1}}
\newcommand*{\Mpch}{\,\text{Mpc}\,h^{-1}}

\newcommand*{\Msunh}{\,M_\odot h^{-1}}

\renewcommand{\added}[1]{\textcolor{explain}{#1}}
\renewcommand{\deleted}[1]{\textcolor{explain}{\sout{#1}}}

\renewcommand{\added}[1]{#1}
\renewcommand{\deleted}[1]{}

\hypersetup{pdfauthor={Lehmann, Mao, Becker, Skillman \& Wechsler},pdftitle={The Concentration Dependence of the Galaxy--Halo Connection}}

\shorttitle{Assembly Bias \& Abundance Matching}
\shortauthors{Lehmann, Mao, Becker, Skillman \& Wechsler}

\begin{document}
\title{The Concentration Dependence of the Galaxy--Halo Connection: \\
  Modeling Assembly Bias with Abundance Matching}
\author{Benjamin~V.~Lehmann}
\affiliation{Kavli Institute for Particle Astrophysics and Cosmology, Department of Physics, Stanford University, Stanford, CA 94305, USA}
\affiliation{SLAC National Accelerator Laboratory, Menlo Park, CA, 94025, USA}
\affiliation{Department of Physics, University of California, Santa Cruz, CA 95064, USA}

\author{Yao-Yuan~Mao}
\affiliation{Kavli Institute for Particle Astrophysics and Cosmology, Department of Physics, Stanford University, Stanford, CA 94305, USA}
\affiliation{SLAC National Accelerator Laboratory, Menlo Park, CA, 94025, USA}
\affiliation{Department of Physics and Astronomy and Pittsburgh Particle Physics, Astrophysics, and Cosmology Center (PITT PACC), University of Pittsburgh, Pittsburgh, PA 15260, USA}

\author{Matthew~R.~Becker}
\affiliation{Kavli Institute for Particle Astrophysics and Cosmology, Department of Physics, Stanford University, Stanford, CA 94305, USA}
\affiliation{SLAC National Accelerator Laboratory, Menlo Park, CA, 94025, USA}
\affiliation{Civis Analytics, West Loop, Chicago, IL 60607, USA}

\author{Samuel~W.~Skillman}
\affiliation{Kavli Institute for Particle Astrophysics and Cosmology, Department of Physics, Stanford University, Stanford, CA 94305, USA}
\affiliation{SLAC National Accelerator Laboratory, Menlo Park, CA, 94025, USA}
\affiliation{Descartes Labs, 1350 Central Avenue, Los Alamos, NM 87544, USA}

\author{Risa~H.~Wechsler}
\affiliation{Kavli Institute for Particle Astrophysics and Cosmology, Department of Physics, Stanford University, Stanford, CA 94305, USA}
\affiliation{SLAC National Accelerator Laboratory, Menlo Park, CA, 94025, USA}

\begin{abstract}
Empirical methods for connecting galaxies to their dark matter halos have become essential for interpreting measurements of the spatial statistics of galaxies. 
In this work, we present a novel approach for parameterizing the degree of concentration dependence in the abundance matching method.
This new parameterization provides a smooth interpolation between two commonly used matching proxies: the peak halo mass and the peak halo maximal circular velocity. This parameterization controls the amount of dependence of galaxy luminosity on halo concentration at a fixed halo mass.
Effectively this interpolation scheme enables abundance matching models to have adjustable assembly bias in the resulting galaxy catalogs.
With the new $400 \Mpch$ DarkSky Simulation, whose larger volume provides lower sample variance, we further show that low-redshift two-point clustering and satellite fraction measurements from SDSS can already provide a joint constraint on this concentration dependence and the scatter within the abundance matching framework.
\end{abstract}

\keywords{dark matter ---  galaxies: halos --- methods: analytical ---
  methods: numerical}

\section{Introduction} \label{sec:intro}

Understanding the connection between galaxies and their dark matter halos is at the heart of modern cosmology and astrophysics. Galaxies are our primary tool to probe the spatial distribution of dark matter and its evolution, both of which are being mapped at increasingly high precision with cosmological surveys \added{(see, e.g., the Sloan Digital Sky Survey, the Dark Energy Survey, the Dark Energy Camera Legacy Survey, and the Large Synoptic Survey Telescope)}.  However, because galaxies are biased tracers of this distribution, taking full advantage of these measurements requires accurate and flexible models for modeling the connection between galaxies and their dark matter halos.
In addition, understanding the statistical mapping between galaxies and halos provides key insights into the physical processes responsible for galaxy formation.

The effects of assembly bias, in particular, remain a significant uncertainty in modeling the galaxy--halo connection \citep{Zentner2014}. 
In dark-matter-only cosmological simulations, it has been shown that halo concentration, along with other properties of the halos and their assembly histories, can have an impact on halo clustering, generally known as halo assembly bias \citep[e.g.,][]{Wechsler2001,Gao2005,Wechsler2006,Croton2007}. Despite a series of studies on the \added{possible} observational evidence for assembly bias \citep{Yang2006,Tinker2012,Lin2015,Miyatake2015,More2016}, the extent to which halo assembly bias results in observable bias in the galaxy population remains highly uncertain. 

Thus it is critical to characterize the assembly bias inherited through the galaxy--halo connection. 
For hydrodynamic simulations and semi-analytic models (SAMs), galaxy assembly bias is an end product rather than a controlled parameter, as these two methods attempt to incorporate the microscopic physics of galaxy formation. (For the latest large-scale hydrodynamic simulations, see, e.g., \citealt{Vogelsberger2014,2015MNRAS.450.1937C,2015MNRAS.446..521S}; for SAMs, see e.g., \citealt{Bower2006,Croton2006,Somerville2008}\added{; see also \citealt{2015ARA&A..53...51S} for a review of galaxy formation models.})
In principle, one can directly characterize the assembly bias for each set of parameter values used in these methods. 
Practically, hydrodynamic simulations are computationally expensive, even when used to produce a handful of realizations. With SAMs, while it is possible to generate many different realizations, the large number of parameters makes it challenging \citep[though not impossible, see, e.g.,][]{Lu2014,Henriques2015} to explore and statistically constrain the full parameter space.
Also, neither hydrodynamic simulations nor SAMs have been shown to reproduce the detailed clustering properties of observed galaxies at the accuracy with which they have been measured, partly due to our incomplete understanding of star formation and feedback mechanisms.

On the other hand, the widely used, conventional halo occupation distribution (HOD) models prescribe the probability that a halo of a given mass $M$ hosts $N$ galaxies above a given luminosity threshold, $P(N|M)$, commonly with a parameterized functional form \citep{Peacock2000,Seljak2000,Scoccimarro2001,Berlind2002,Cooray2002,Bullock2002}. In this fashion, the HOD approach erases much of the halo assembly bias, as it explicitly assumes that the galaxy population in a halo depends only on its mass. Recently, some HOD models incorporate dependence on other parameters \citep[e.g.,][]{Paranjape2015,Hearin2016}. In particular, \citet{Hearin2016} parameterize the assembly bias in an HOD-like model \added{ (``Decorated HOD''), and later \citet{1606.07817} and \citet{1610:01991} further constrain the assembly bias parameters in the Decorated HOD model with SDSS data. The essence of their work is closely related to this work, but with the HOD framework, which makes a different set of assumptions than we do here.}

In this work, we characterize the assembly bias in another commonly used empirical model of the galaxy--halo connection: the abundance matching technique (or subhalo abundance matching, SHAM). 
Abundance matching is a fairly generic scheme for linking galaxies with dark matter halos, without a full description of baryonic physics \citep{Kravtsov2004,Vale2004,Vale2006,Conroy2006}.
The basic assumption of abundance matching is that galaxies live in halos, and one particular galaxy property (typically luminosity or stellar mass) is approximately monotonically related to a halo property (typically virial mass, $\sub{M}{vir}$, or maximum circular velocity, $\sub{v}{max}$), by matching their ``abundance'' (i.e., matching at fixed number densities). 

A major strength of abundance matching is the fact that it uses the full predictive power of the cosmological model, including the predictions for the number and properties of substructures and their relation to their host halos.  Certain abundance matching models have been shown to reproduce the observed two-point correlation function with surprising accuracy, with only a very small number of parameters \citep{Conroy2006, Trujillo-Gomez2011,Reddick2013}, as well as three-point statistics, galaxy--galaxy lensing, and the Tully--Fisher relation \citep[e.g.,][]{Marin2008,Tasitsiomi2004,Desmond2015}.
Similar models have also been shown to reproduce a wide range of other statistics of the galaxy distribution \citep{Hearin2013a,Hearin2014}.

The abundance matching parameters that have typically been considered are the \emph{scatter} in the galaxy--halo relation, usually in terms of the standard deviation of the galaxy luminosities or stellar masses at a fixed value of the halo property, and the choice of halo property.  Commonly used halo properties (or \emph{proxies}) include the halo mass ($\sub{M}{vir}$ or variants), the maximum circular velocity $\sub{v}{max}$, and these two properties evaluated at different epochs.  For example, \cite{Reddick2013} perform a systematic search for a best-fit model to spatial clustering and the conditional luminosity function and find that using the peak value of $\sub{v}{max}$ throughout all timesteps (i.e., $\sub{v}{peak}$) as the proxy with a scatter of $\sim$~0.2 dex gives the best predictions.  Other studies obtain similar results \citep[e.g.,][]{Chaves-Montero2015,Guo2015}.

Although different proxies have different physical meanings attached to them, abundance matching is only sensitive to the relative rankings of halos when they are ranked by the proxy in consideration. 
Hence, the seemingly distinct choices of using proxies based on $\sub{v}{max}$ or $\sub{M}{vir}$ are merely different ways to rank the halos. 
For instance, ranking halos by $\sub{v}{max}$ is similar to ranking by $\sub{M}{vir}$, except that more concentrated halos are given a higher rank, since at a fixed $\sub{M}{vir}$, more concentrated halos have larger $\sub{v}{max}$ \citep{Klypin2011}.
As a result, this choice influences the dependence of galaxy luminosity or stellar mass on halo concentration at a given halo mass. 
Our current understanding of galaxy formation physics is not yet sophisticated enough to quantify this concentration dependence, and hence it is natural to parametrize this dependence on concentration by continuously interpolating the rankings that different proxies give. 
Furthermore, such a parametrization also provides a natural way to control how halo assembly bias propagates to galaxy assembly bias in an observed population.

This work is the first to present results on the clustering statistics using abundance matching with a continuous parameter controlling the matching proxy, and hence the amount of concentration dependence and assembly bias.
This work is also \added{one of} the first \added{studies} to compare the observed two-point clustering with a cosmological box of $(400\Mpch{})^3$ at a mass resolution of better than $\sim 10^8 \Msunh$ (from the ``Dark Sky'' Simulations).  The large volume of this box yields much tighter constraints on abundance matching parameters, which provide further insight into the amount of galaxy assembly bias present. \added{(See also \citealt{Guo2015,1608.03660} for the study of spatial correlation functions using the ``SMDPL'' simulation, which has similar volume and resolution as this Dark Sky $400\Mpch{}$ box.)}

Note that this work differs from the recent development on the two-parameter abundance matching technique (commonly known as conditional abundance matching, CAM), which attempts to match two halo proxies with two galaxy properties \citep{Hearin2014,Kulier2015}. The model we propose in this work, by contrast, still matches one halo proxy with one galaxy property, yet the halo proxy in consideration is a linear combination of two different halo properties.
The proposed technique to combine distinct halo properties into one matching proxy can still apply to other variants of abundance matching, including CAM.

This paper is organized as follows. 
In \autoref{sec:data} we describe the simulations and the observed catalogs used in this study, 
and also describe the procedure for generating mock galaxy catalogs and the covariance. 
In \autoref{sec:model} we present this novel model of concentration-dependent abundance matching and explore how the new parameter affects the galaxy clustering, the satellite fraction, and the assembly bias. 
In \autoref{sec:constraints} we compare the galaxy clustering signals from this model and from observations to constrain the model parameters. 
We then discuss some detailed aspects of our results in \autoref{sec:discussion}, and summarize this paper in \autoref{sec:summary}.

\section{Simulations and Galaxy Catalogs} \label{sec:data}

\subsection{Simulations}

\begin{deluxetable*}{r c l c c c c l l}
    \tablewidth{\textwidth}
    \tablecaption{Cosmological and Simulation Parameters for Boxes Used in This Study \label{tab:cosmosim}}
    \tablehead{
      \colhead{Box Name} & \colhead{Side Length} & \colhead{Particle} & \colhead{$h$} & \colhead{$\Omega_M$} & \colhead{$n_s$}  & \colhead{$\sigma_8$} & \colhead{Particle Mass} & \colhead{Code} \\
      & [$\Mpch$] & number & & & & & [$\Msunh$] & }
    \startdata
    \texttt{c250-2048}    &  \phn250 &  \phn2048$^3$ & 0.7\phn\phn & 0.286 & 0.96\phn & 0.82\phn & $1.44\times 10^8$ & \textsc{L-Gadget}   \\
    \texttt{Bolshoi}      &  \phn250 &  \phn2048$^3$ & 0.7\phn\phn & 0.27\phn  & 0.95\phn & 0.82\phn & $1.35\times 10^8$ & \textsc{Art}        \\
    \texttt{BolshoiP}    &  \phn250 &  \phn2048$^3$ & 0.678 & 0.295 & 0.968 & 0.823 & $1.49\times 10^8$ & \textsc{Art}     \\
    \texttt{MDPL}         & 1000 &  \phn3840$^3$ & 0.678 & 0.307 & 0.96\phn  & 0.823 & $1.51\times 10^9$ & \textsc{L-Gadget} \\
    \texttt{DarkSky-250}  &  \phn250 &  \phn2560$^3$ & 0.688 & 0.295 & 0.968 & 0.834 & $7.63\times 10^7$ & \textsc{2Hot}     \\
    \texttt{DarkSky-400}  &  \phn400 &  \phn4096$^3$ & 0.688 & 0.295 & 0.968 & 0.834 & $7.63\times 10^7$ & \textsc{2Hot}     \\
    \texttt{DarkSky-Gpc} & 1000 &  10240$^3$~\tablenotemark{a} & 0.688 & 0.295 & 0.968 & 0.834 & $4.88\times 10^9$~\tablenotemark{b} & \textsc{2Hot} \\
    \enddata
\tablenotetext{a}{Halo catalogs and merger trees are constructed with
1/32 of the total particle number.}
\tablenotetext{b}{Effective mass of the down-sampled particles.}
\end{deluxetable*}

This study uses several cosmological boxes, as listed in \autoref{tab:cosmosim}.
The \texttt{c250-2048} box comes from the ``Chinchilla'' suite, run with the \textsc{L-Gadget} code, a variant of \textsc{Gadget} \citep{Springel2005}. The ``Chinchilla'' suite spans a wide range of box sizes and resolutions with the same cosmology (M.~R.~Becker et al.\ 2016, in preparation; some boxes in this series were described in \citealt{Mao2015} and were also used in \citealt{Desmond2015}). 
\texttt{Bolshoi} and \texttt{BolshoiP} have the same box size and resolution as \texttt{c250-2048}, but have different cosmologies and were run with the ART $N$-body code \citep{Klypin2011}. \texttt{MDPL} is part of the ``MultiDark'' suite \citep{Klypin2014}, and was also run with \textsc{Gadget}.  The three \texttt{DarkSky} boxes of different sizes are smaller boxes that accompany the $8 \Gpch{}$ box from the ``Dark Sky'' Simulations \citep{Skillman2014}, run with the \textsc{2HOT} code \citep{Warren2013}.  Here we refer to these boxes as {\tt DarkSky-250 (ds14\_j\_2560)}, {\tt DarkSky-400 (ds14\_i\_4096)}, and {\tt DarkSky-Gpc (ds14\_b)}. The particles used to build the halo catalogs and merger trees for the  \texttt{DarkSky-Gpc} box were down-sampled (1/32 particles) from a high-resolution box run with $10240^3$ particles.

For each of these boxes, we use the halo catalog generated by the \textsc{Rockstar} halo finder \citep{Behroozi2013a} and the \textsc{Consistent Trees} merger tree builder \citep{Behroozi2013b}.
We use the virial overdensity ($\sub{\Delta}{vir}$) as our halo mass definition \citep{Bryan1998}. 

\subsection{Mock Galaxy Catalogs}

The mock galaxy catalogs used in this work are generated with the abundance matching technique. We follow the procedure of \cite{Behroozi2010} and \cite{Reddick2013} in order to implement abundance matching with scatter in luminosity at fixed halo proxy. First, we deconvolve the scatter from the luminosity function. We then abundance match luminosity with the halo proxy, producing a catalog of galaxy luminosities. Finally, we replace the scatter by adding a log-normal scatter to the catalog.

We make measurements of the projected two-point correlation function, $\wprp$, from the mock catalogs as follows. We use the plane-parallel approximation in redshift-space and integrate along one of the axes (i.e., the line of sight), with an integration depth of $2\sub{z}{max} = 80 \Mpch$. Redshift-space distortions are applied along the integration axis before integration. We account for the periodic boundary conditions of the cosmological boxes when computing the projected correlation function, \added{and hence we can evaluate the expected number of uniformly random pairs explicitly without using an estimator.}

\subsection{SDSS Galaxy Catalogs}

In this study, we use the luminosity function (for abundance matching) and the two-point clustering measurements (for comparison) extracted by \cite{Reddick2013}.  These measurements were made on the volume-limited samples from the New York University Value Added Galaxy Catalog (NYU-VAGC;  \citealt{Blanton2005}), based on Data Release 7 from the Sloan Digital Sky Survey \citep{Padmanabhan2008,Abazajian2009}. We note that these measurements are quite consistent with the measurements of \cite{Zehavi2011}, but here a consistent sample was used to determine both the luminosity function and clustering measurements. We refer the readers to Section 2 and Appendix C of \cite{Reddick2013} for details on these measurements.  In this work, we focus primarily on constraining our models with 
galaxies of luminosity $\sim L_*$ and brighter in order to be conservative about the resolution requirements for 
the complete halo and subhalo samples needed for abundance matching, but we present results from dimmer samples 
in \autoref{sec:dimmer}.

\subsection{Calculating the Covariance}

In \autoref{sec:constraints}, when we compare the SDSS data to the predicted $\wprp$ obtained from the mock catalogs with a $\chi^2$ statistic, \added{we need to take the covariance matrix into account. The covariance matrix has three contributions: (1) the sample variance (jackknife covariance) of the SDSS data, (2) the sample variance (jackknife covariance) of the mock catalog, and (3) the covariance due to the stochastic scatter in the mock catalog.
We sum up these three contributions quadratically to obtain the final covariance matrix.}

First, the sample variance in the SDSS measurements is estimated by jackknifing the SDSS data set, as detailed in \cite{Reddick2013}.
Second, we estimate the sample variance in the predictions of abundance matching due to the finite volume of the $N$-body simulations 
used in this work. We employ a jackknife procedure in order to estimate the contribution to the covariance matrix from this effect. Each $N$-body box is divided into smaller square ``sky areas.'' \added{Each patch has side lengths of $25\Mpch$, $25\Mpch$, and the original box side length}. We then omit one patch at a time in the jackknifing process (i.e., omitting everything along the line of sight in the square patches), and compute the jackknife covariance.  Thirdly, the final contribution to the covariance comes from the scatter in abundance matching. Since we apply log-normal random scatter in luminosity directly to the catalogs, multiple catalogs generated with the same abundance matching parameters produce slightly different predictions for $\wprp$. Thus, from each set of abundance matching parameters, we generate 40 catalogs, compute $\wprp$ for each, and calculate the covariance on $\wprp$ due to this random variation.

We note that the estimate of the covariance of the mock $\wprp$ has a direct impact on the goodness of fit, and hence on the derived constraints on the abundance matching parameters. Nevertheless, \cite{Norberg2009} find that the jackknife method does not typically underestimate the covariance. 

\section{Abundance Matching with Adjustable Concentration Dependence} \label{sec:model}

\subsection{Interpolating between Abundance Matching Proxies}
\label{sec:model-alpha}

\begin{figure}[htb!]
  \putfig{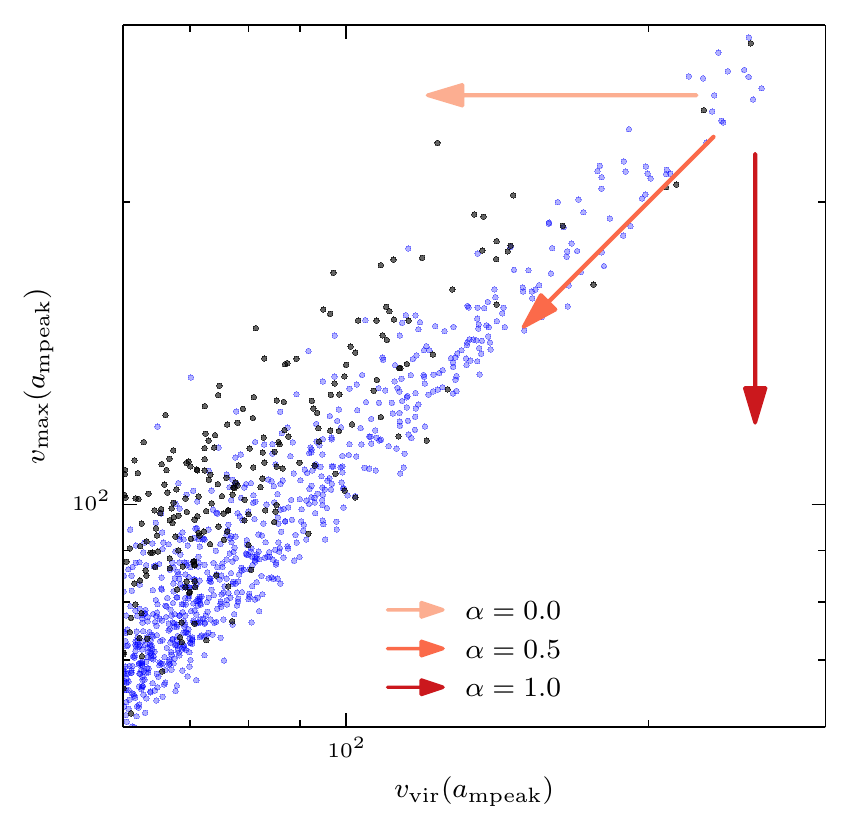}
  \caption{Relation between the two halo properties $\sub{v}{vir}$ and $\sub{v}{max}$ (both evaluated at $\sub{a}{mpeak}$) and abundance matching rankings.
  Each point represents a host halo (blue) or a subhalo (black). The total number of halos is down-sampled for illustration purposes.
   Each arrow shows the direction of decreasing abundance matching rank when a particular value of $\alpha$ is used (from light to dark: $v_{\alpha=0}=\sub{v}{vir}$, $v_{\alpha=0.5}$, and $v_{\alpha=1}=\sub{v}{max}$). The figure indicates how the choice of proxy impacts both the fraction of subhalos that are included in the sample, as well as the concentration of the included halos, which will impact their clustering properties.}
  \label{fig:generalized-proxy}
\end{figure}

\added{A key simplifying assumption in the abundance matching framework is that some galaxy property (for both centrals and satellites) is tightly correlated with some dark matter (sub)halo property.  Physically, this halo property should be connected to, for example, the mass, velocity, or potential of the dark matter halo. However, at present, we do not understand the details of galaxy formation well enough to specify the exact connection. Hence, in practice, the halo proxy used in abundance matching could be a function of multiple halo properties instead of a single halo property.}

\added{In particular, to test whether, or how strongly, galaxy properties within halos at a fixed mass should be dependent on halo concentration or assembly history,} here we present an interpolation scheme that generalizes the conventional abundance matching model to allow continuously adjustable concentration dependence.
To build such a scheme, we adopt the parameterization used in \cite{Mao2015}, defining a new generalized proxy to be used in abundance matching:
\begin{equation}
\label{eq:def_proxy}
v_\alpha :=
\sub{v}{vir} \left( \frac{\sub{v}{max}}{\sub{v}{vir}}  \right)^\alpha,
\end{equation} 
where $\sub{v}{max}$ is the maximal circular velocity and 
\begin{equation} \label{eq:vvir}
\sub{v}{vir} :=  \left( \frac{G \sub{M}{vir}}{\sub{R}{vir}}\right)^{1/2} = 
\left( \frac{4\pi}{3} \sub{\Delta}{vir} \sub{\rho}{crit} G^3 \right)^{1/6} \sub{M}{vir}^{1/3}, 
\end{equation}
with $\sub{\Delta}{vir}$ being the virial overdensity and $\sub{\rho}{crit}$ the critical density. 

This generalized proxy captures the continuously varying dependence on concentration through the parameter $\alpha$ because the ratio $\sub{v}{max} / \sub{v}{vir}$ can be viewed as a proxy for halo concentration.
In principle, this ratio can be replaced by $f(c)$ with a general function $f$. 
Nevertheless, using this ratio facilitates comparisons with other proxies that have been used in the literature. In particular, when $\alpha = 0$, the dependence on concentration is turned off, as $v_{\alpha=0} = \sub{v}{vir} \propto \sub{M}{vir}^{1/3}$, and when $\alpha = 1$, this proxy reduces to the maximal circular velocity $v_{\alpha=1} = \sub{v}{max} $.

\added{Note that Equation~\eqref{eq:def_proxy} can be written as 
\begin{equation}
\log v_\alpha = \alpha \log \sub{v}{max} + (1-\alpha) \log \sub{v}{vir}.
\end{equation}
Hence, this new proxy can be considered as a linear combination of $ \log \sub{v}{max}$ and $ \log \sub{v}{vir}$ on a logarithmic scale. As illustrated in \autoref{fig:generalized-proxy}, on a log--log plot of $\sub{v}{max}$ and $\sub{v}{vir}$, the value of $\alpha$ affects the direction of the halo ranking.}
In this figure, the arrow represents the direction of descending rank when the halos are ranked by $v_\alpha$, and the slope of the arrow is $\alpha/(1-\alpha)$. Hence, different values of $\alpha$ effectively rank the halos with different slopes. As a result, at a given number density, different values of $\alpha$ select out different halos. In particular, a larger value of $\alpha$ selects out more low-mass, high-concentration halos, and also more subhalos.

We note that the specific choice of the parameterization of the concentration dependence should not impact our results significantly, as the essence of our model is to vary how much we weight the concentration of halos when ranking halos by their masses in the abundance matching procedure. However, one could instead weight other halo properties, such as halo formation time, in order to study the dependence on other properties in abundance matching. In this work, we only study the dependence on concentration.  Nevertheless, we expect that qualitatively similar results would also apply to other proxies that are highly correlated with concentration.

\subsection{Evaluating the Proxy at the Epoch of Peak Mass}

So far, we have only discussed how to model the concentration dependence in our new proxy. In abundance matching, the choice of epoch at which the ranking proxy is evaluated also significantly impacts the results \citep{Reddick2013,Chaves-Montero2015}. For example, if the proxy is evaluated at the present day, subhalos are usually ranked lower due to stripping, and the resulting mock catalog is less clustered. \cite{Conroy2006} argues that the time at which a subhalo enters the virial radius of its parent halo is a natural time at which to set proxies.
\cite{Reddick2013} further uses the peak values of those proxies (e.g., $\sub{M}{peak}$ and $\sub{v}{peak}$) throughout history. 
In this work, we limit our discussion to a single choice of epoch. We evaluate the value of $v_\alpha$ for each halo at the epoch when $\sub{M}{vir}$ reaches its peak value, and let $\hat{v}_\alpha$ denote this quantity.  In follow-up work, we will explore this choice of proxy epoch in detail.

Since $\hat{v}_\alpha$ is evaluated at the time of peak mass for each halo, ranking with $\hat{v}_{\alpha=0}$ and $\hat{v}_{\alpha=1}$ is equivalent to ranking with $\sub{M}{peak}$ and $\sub{v}{max}$ at $\sub{M}{peak}$ respectively. (However, the former is only approximately true in our case because $\sub{\Delta}{vir}$ in Equation~\eqref{eq:vvir} has a weak dependence on the scale factor, and for different halos, $\sub{M}{peak}$ occurs at different scale factors.)  Our choice of evaluating the abundance matching proxy at the scale when $\sub{M}{peak}$ rather than when $\sub{v}{peak}$ occurs was motivated by the finding that halos at the largest circular velocities may be out of dynamical equilibrium \citep{Ludlow2012}; e.g. \cite{Behroozi2014} showed that $\sub{v}{peak}$ is commonly set by major mergers, and hence may not represent the physical time at which the subhalo started to be stripped.  Evaluating the proxy at the scale factor of $\sub{M}{peak}$ then avoids this unphysical epoch probed by $\sub{v}{peak}$, and is similar to using the relaxation criterion proposed by \cite{Chaves-Montero2015}.
Nevertheless, for the purpose of abundance matching, the difference between matching to $\sub{v}{peak}$ and to $\sub{v}{max}(\sub{a}{mpeak})$ is minimal, as the rank orders are very similar when halos are ranked by these two proxies. As a result, the clustering signals with these two proxies are also similar.

\subsection{Impact of \texorpdfstring{$\alpha$}{alpha} on Clustering}
\label{sec:results-clustering}
  
\begin{figure*}[htb!]  
\putfig{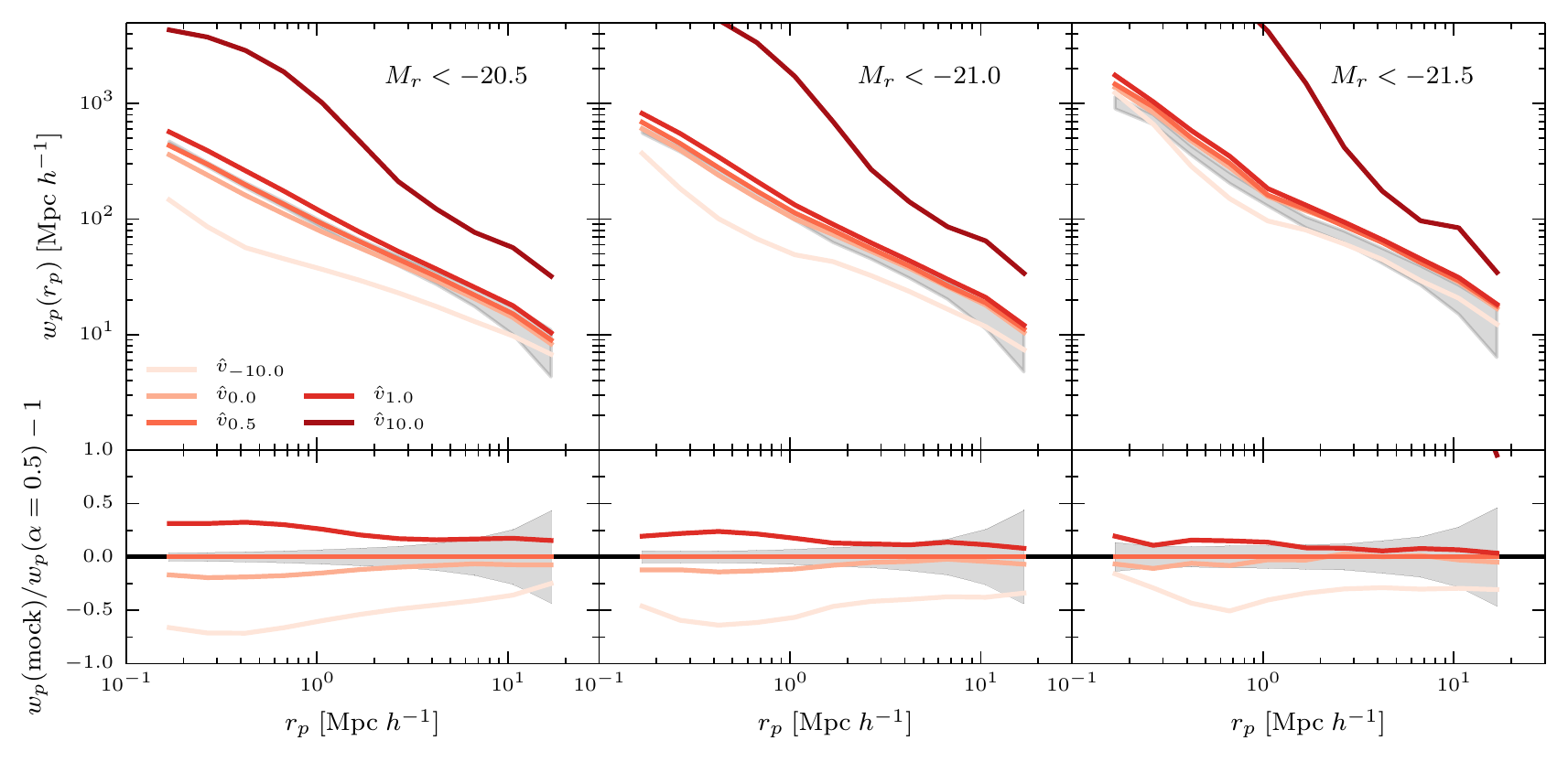}
\caption{Dependence of galaxy clustering on the abundance matching proxy.
Top row shows $w_p(r_p)$ for three thresholds ($M_r < -20.5$, $-21$, and $-21.5$; from left to right) in the \texttt{DarkSky-400} box. 
Lines of different colors show different values of $\alpha$ ($-10$, $0$, $0.5$, $1$, $10$; from light to dark). 
Larger values of $\alpha$ correspond to stronger concentration dependence.
The gray band shows the SDSS measurements and the errors combined with mock errors. 
Bottom row shows the relative difference in $w_p(r_p)$ with respect to $\hat{v}_{\alpha=0.5}$.
}
\label{fig:wps-selected-alpha}
\end{figure*}

\begin{figure*}[htb!]
\putfig{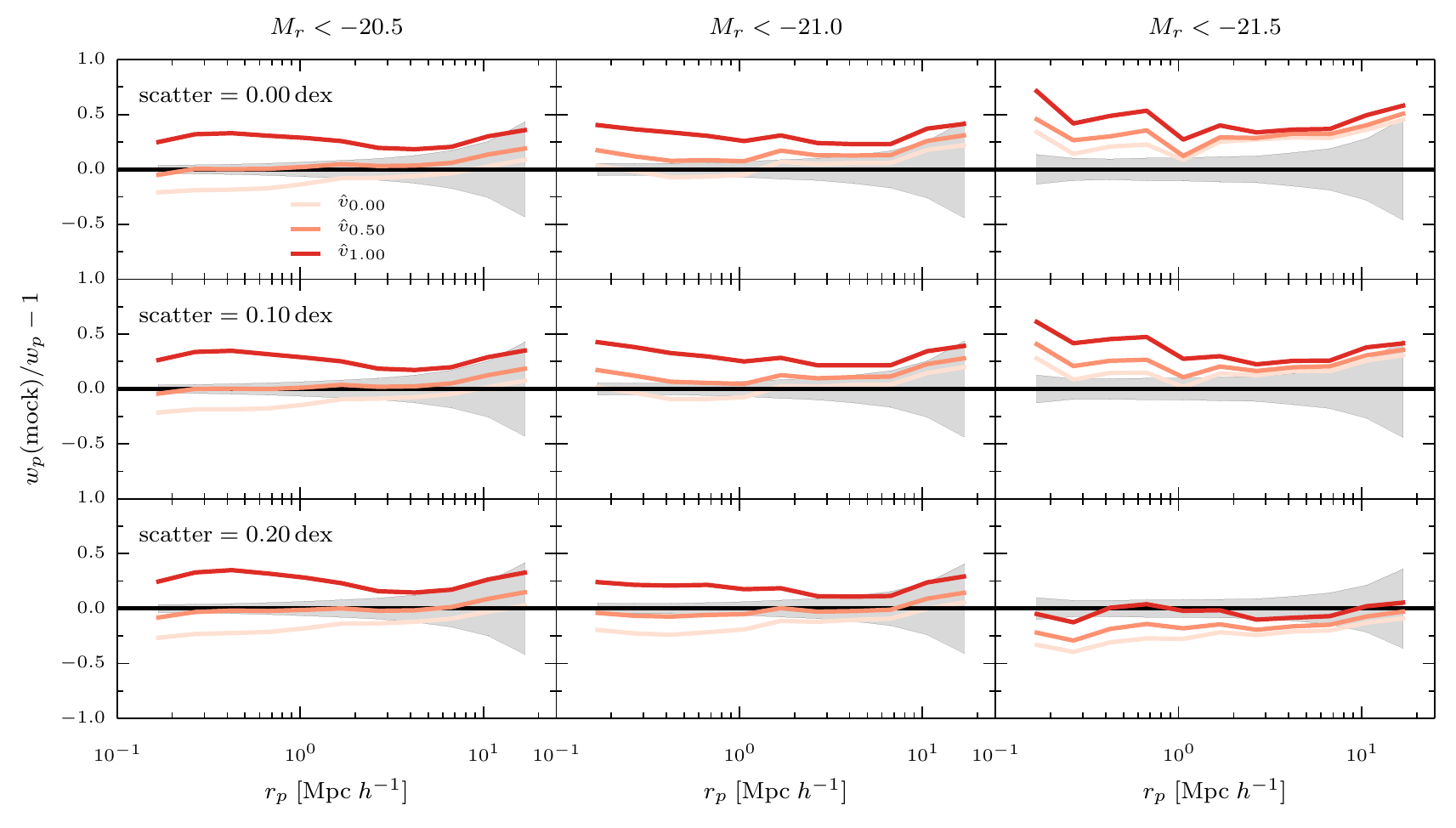}
\caption{Dependence of galaxy clustering on the abundance matching scatter and proxy.
Relative difference in $w_p(r_p)$ between the \texttt{DarkSky-400} galaxy catalog and the SDSS measurements, for three thresholds ($M_r < -20.5$, $-21$, and $-21.5$; from left to right), three values of scatter (0, 0.15, and 0.25; from top to bottom), and three values of $\alpha$ (0, 0.5, and 1; from light to dark). 
Gray bands indicate combined SDSS and mock errors.  
}
\label{fig:wps-x-scatter}
\end{figure*}

We first demonstrate the impact of $\alpha$, as defined in \autoref{sec:model-alpha}, on clustering. 
\autoref{fig:wps-selected-alpha} shows the wide range of clustering predictions that can be produced by varying $\alpha$. We find that changing $\alpha$ can significantly change the clustering, and that a higher value of $\alpha$ produces a more clustered catalog. 

There are two effects that contribute to this result.  First, at a given halo mass, on average, subhalos have higher concentrations than host halos. Hence, when a higher value of $\alpha$ is used, subhalos are more likely to make it through the threshold cut, resulting in a more clustered sample. Effectively, increasing $\alpha$ increases the difference between the luminosity--halo mass relation of host halos and that of subhalos. This effect impacts both the one- and two-halo terms, and also boosts the satellite fraction.

Second, when we use a higher value of $\alpha$, high-concentration halos are ranked higher in the catalog and are more likely to make it through the threshold cut.  Since, in this mass regime, high-concentration halos are more clustered due to halo assembly bias \citep{Wechsler2006}, the resulting catalog is also more clustered. This effect impacts mostly the two-halo term, and is less significant in brighter samples. In the sections below, we discuss these two effects in detail.

It is known that increasing the scatter in abundance matching would decrease clustering strength because it brings in lower-mass halos \citep[e.g.,][]{Reddick2013}. Thus, there exists a degeneracy between $\alpha$ and the scatter. This degeneracy is demonstrated in \autoref{fig:wps-x-scatter}, which shows the correlation function for several values of $\alpha$ and scatter. The clustering strength decreases with decreasing $\alpha$ and also with increasing scatter. 
Nevertheless, the scatter has a stronger effect on the brighter samples, while $\alpha$ has a stronger effect on the fainter samples. This implies that samples of different thresholds are likely to give different constraints on $\alpha$ and scatter, and might be able to break the degeneracy between $\alpha$ and scatter.

\subsection{Impact of \texorpdfstring{$\alpha$}{alpha} on the Satellite Fraction} \label{sec:discussion-fsat}

\begin{figure}[htb!]
  \putfig{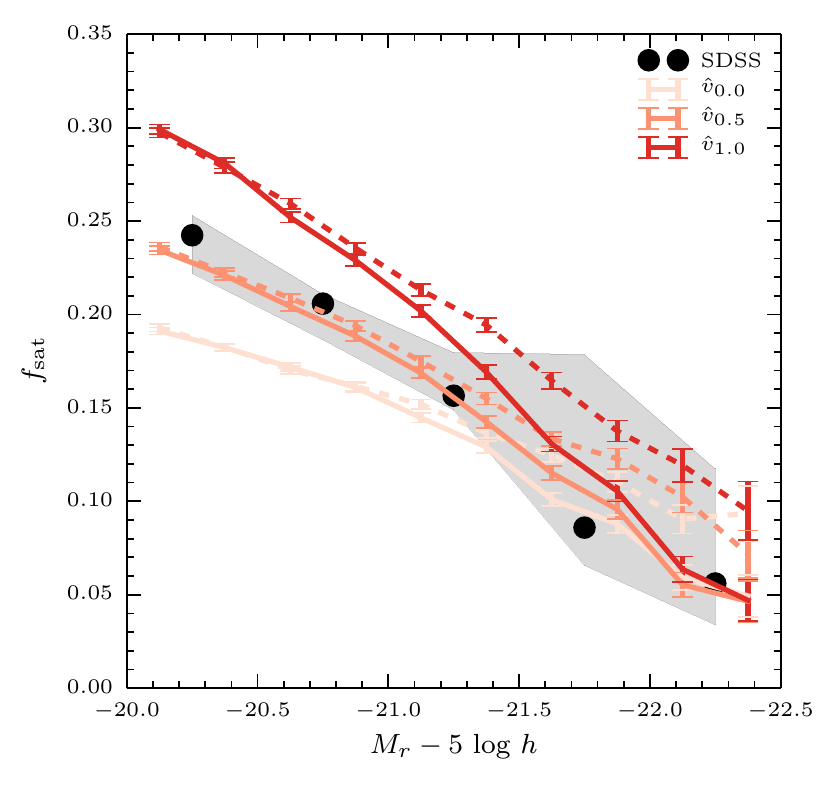}
  \caption{Satellite fraction as a function of luminosity, for three values of $\alpha$ (0, 0.5, and 1.0; from light to dark), computed with
    zero scatter (solid) and 0.15-dex scatter (dashed), using the \texttt{DarkSky-400} box. Error bars show the jackknifing error. Circles show the satellite fraction measured from SDSS groups \citep{Reddick2013},
    and the gray band indicates the sum of the error from SDSS data and the 
    estimated systematic error introduced by the group finder (see the text of \autoref{sec:discussion-fsat} for details).}
  \label{fig:fsat-vs-alpha}
\end{figure}

\begin{figure*}[htb!]
  \putfig{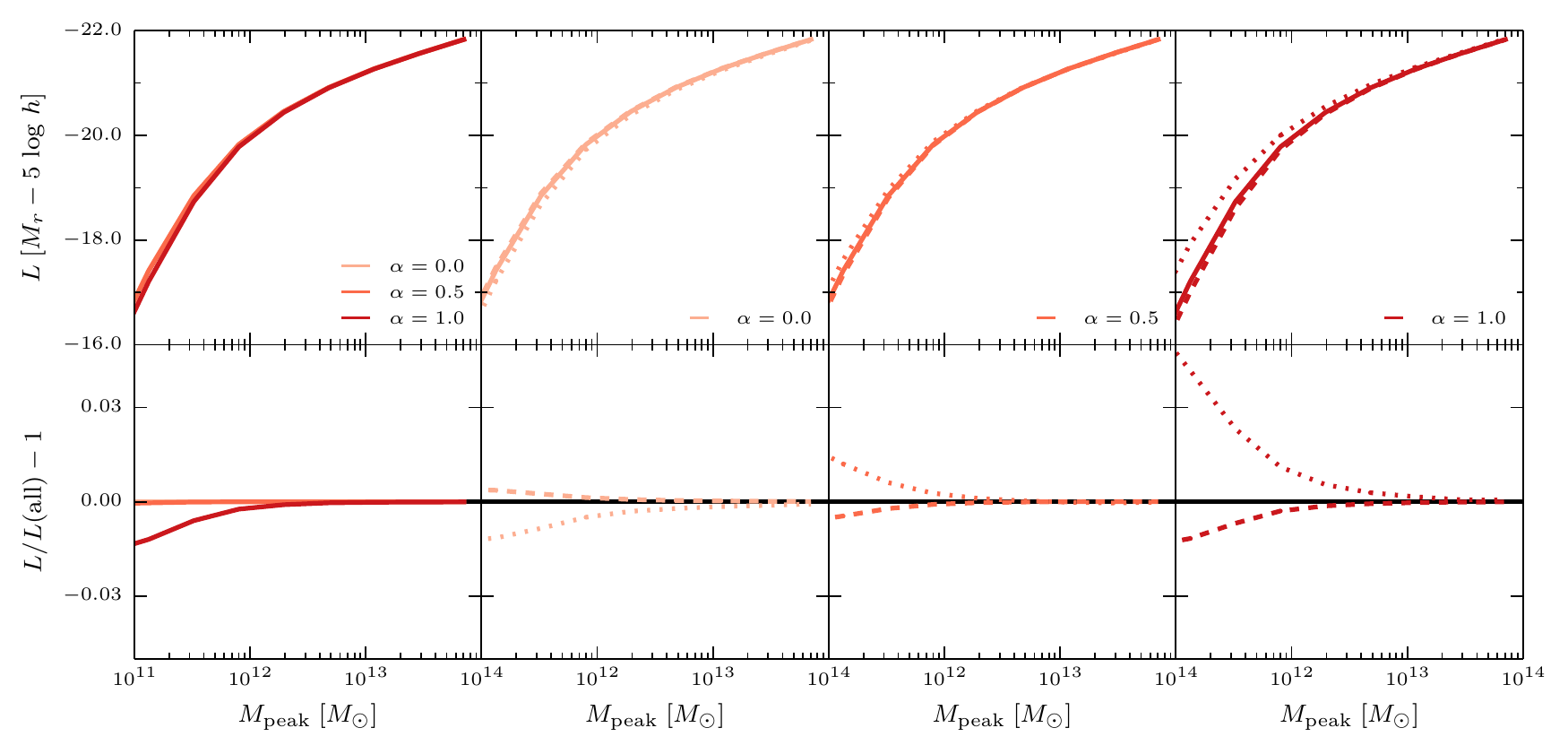}
  \caption{Luminosity--halo mass relation for several values of $\alpha$ (0, 0.5, and 1.0; from light to dark).
      Lines with different styles show the relation for all halos (solid), host halos only (dashed), and subhalos (dotted) only.
      The leftmost top panel shows that the value of $\alpha$ has very little effect on the relation that includes all halos. The right three panels show that the difference between the relations of central halos and subhalos increases with $\alpha$. 
      The bottom row shows the relative difference with respect to the relation for all halos (and for the leftmost bottom row, with respect to the relation for $\alpha=0$).}
  \label{fig:luminosity-halo-mass}
\end{figure*}

Here, we define the satellite fraction to be the fraction of satellites in bins of luminosity.  In this study, we did not apply a group finder on the mock galaxies, so galaxies labeled as satellites are exactly the same as those labeled as subhalos in the initial halo catalog. That is, the satellite fraction we measured here is actually the fraction of galaxies assigned to subhalos in each luminosity bin. A subhalo is defined as any halo whose center falls within another larger halo. We refer to a halo that is not a subhalo as a host halo.

\autoref{fig:fsat-vs-alpha} shows the satellite fraction as a function of luminosity for several values of $\alpha$. As expected, increasing $\alpha$ increases the satellite fraction, since subhalos are, on average, more concentrated than host halos of the same mass; hence, subhalos are ranked higher when $\alpha$ is larger. This is especially true at the faint end because the ratio $\sub{v}{max}/\sub{v}{vir}$ differs more between subhalos and host halos for low-mass halos.

Applying scatter to abundance matching increases the satellite fraction at the bright end because more satellites in the fainter luminosity bins are scattered up to the brighter luminosity bins. Applying scatter does not significantly change the satellite fraction for samples fainter than $M_r = -21$. 

Another way to demonstrate this change in the satellite fraction is to look at the difference between the luminosity--halo mass relation of host halos (central galaxies) and that of subhalos (satellite galaxies).
  Previous studies have explored the case in which the stellar mass--halo mass relations of central and satellite galaxies differ from each other \citep[e.g.,][]{Neistein2011,Rodriguez-Puebla2012,Rodriguez-Puebla2013}.
  Here, using the $\alpha$ parameter, we can evaluate this difference quantitatively.
  \autoref{fig:luminosity-halo-mass} shows the luminosity--halo mass ($L-M_h$) relations for all halos, only host halos, and only subhalos, for different values of $\alpha$. We see that changing $\alpha$ changes the overall $L-M_h$ relations very little, but changes the difference between the halo and subhalo $L-M_h$ relation.
  Specifically, increasing $\alpha$ effectively more strongly differentiates the $L-M_h$ relations for halos and subhalos, while maintaining the overall $L-M_h$ relation.

In \autoref{fig:fsat-vs-alpha}, we also compare our results with the observed satellite fraction. The observed satellite fraction measurements are taken from \cite{Reddick2013}, who used a group finder \citep{Tinker2011} applied to the same sample used to make the clustering measurements. 
Since we did not apply the same group finding procedure on our mock catalogs, this comparison is subject to the systematic errors introduced by the group finder.
The gray band shown in \autoref{fig:fsat-vs-alpha} is the sum of the error from SDSS data and the estimated systematic error introduced by the group finder; the latter was estimated by taking the one-sided difference between the satellite fractions before and after the catalog was processed with a group finder, shown in the left panel of Figure 21 in \cite{Reddick2013}.
We see that these systematic errors increase significantly at the bright end, due to the fact that the group finder does not always select the right galaxy as the central.  However, both scatter and group finding have much smaller impacts at luminosities dimmer than $M_r = -21$, which is also where $\alpha$ has a larger impact.
Up to the systematic errors, the observed satellite fraction agrees well with the model prediction when $\alpha\sim0.5$. We show in \autoref{sec:am-alpha-scatter} that this is also consistent with the inference from galaxy clustering.

\subsection{Impact of \texorpdfstring{$\alpha$}{alpha} on Assembly Bias} \label{sec:discussion-hab}
\begin{figure*}[htb!]
  \putfig{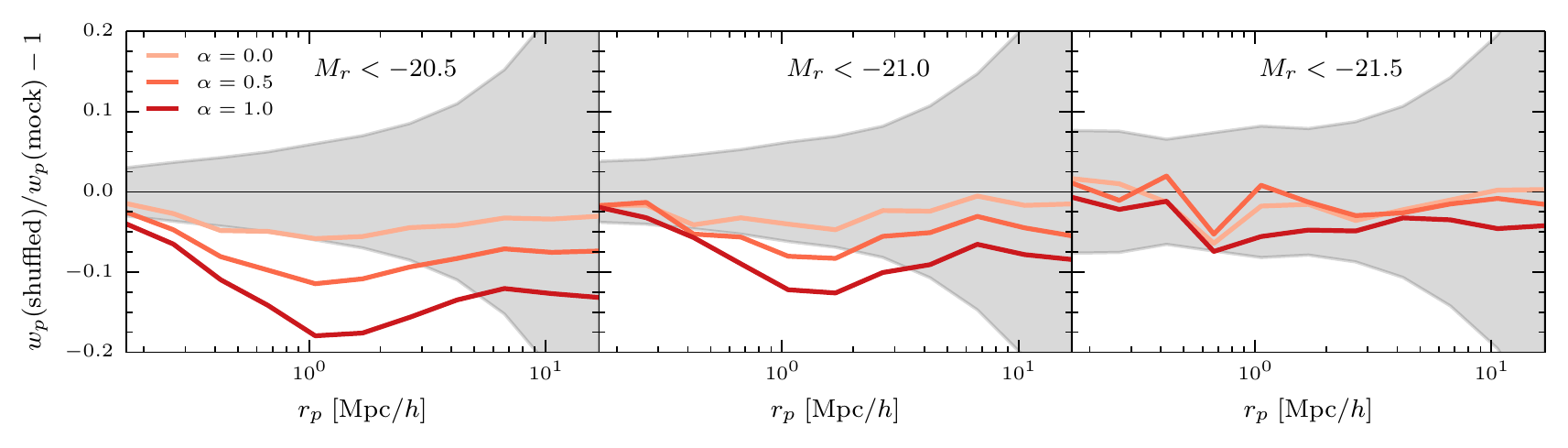}
  \caption{Relative difference in $\wprp$ between shuffled and unshuffled catalogs
    for three thresholds ($M_r<-20.5$, $-21$, and $-21.5$; from left to right) and
    three values of $\alpha$ (-1, 0, 0.5, and 1; from light to dark), for the \texttt{DarkSky-400} box. No scatter is applied in the abundance matching procedure. The gray band shows the combined error from the observed data and the mock catalogs.  Assembly bias increases the large-scale clustering in our best-fit model by $\sim 8\%$ for the dimmest sample shown here.
  }
  \label{fig:wp-shuffled}
\end{figure*}

In our model, $\alpha$ also controls how much halo assembly bias can manifest in the mock catalogs as galaxy assembly bias. To quantify this, we need to separate the effects of the satellite fraction and halo assembly bias.
To that end, we shuffle our mock catalogs to remove halo assembly bias, but leave the satellite fraction intact. Here we adopt the same shuffling procedure as described by \cite{Zentner2014}. We divide the catalogs into bins of halo masses, with a bin width of 0.1 dex. For each bin, we first shuffle the central galaxies, and then independently shuffle the satellites (while retaining their relative positions to the central galaxies). This procedure preserves the halo occupation and the satellite fraction in the catalogs by construction.

\autoref{fig:wp-shuffled} shows the relative difference in $\wprp$ between the  shuffled and unshuffled catalogs. Since the shuffling procedure preserves the satellite fraction, the difference seen in this figure comes from halo assembly bias alone.  We see that the difference is larger for fainter samples and for larger values of $\alpha$. This behavior is expected: halo assembly bias impacts the fainter samples more strongly, and catalogs with a larger value of $\alpha$ have stronger halo assembly bias and are more clustered.  We note here that although models with more concentration dependence have stronger assembly bias, there is still some assembly bias in the models with $\alpha = 0$, because the relationship between $M_\text{vir}$ 
and $M_\text{peak}$ has some dependence on formation time and/or halo concentration.

\autoref{fig:wp-shuffled} also shows the scale dependence of assembly bias for each value of $\alpha$.
We find that halo assembly bias impacts both the two-halo term and the transition regime around $1$--$2\Mpch$, in agreement with the findings of \cite{Sunayama2015}.
At the smallest scales, the original catalog and the shuffled catalog exhibit similar clustering, since the clustering at small scales is dominated by the change in satellite fraction. This implies that our $\hat{v}_\alpha$ model is distinct from merely introducing halo assembly bias to a non-biased catalog (e.g., modeling the HOD). In particular, varying $\alpha$ simultaneously changes the amount of halo assembly bias and the satellite fraction.  

\section{Constraining the Concentration-dependent Model}
\label{sec:constraints}

\subsection{Jointly Constraining \texorpdfstring{$\alpha$}{alpha} and Scatter}

In the previous section, we present how this $\alpha$ parameter, which controls the concentration dependence in abundance matching, affects the clustering signals in the mock catalog. 
Given this finding, here we investigate whether the current galaxy clustering measurement can already provide constraints on the this $\alpha$ parameter.
Since the effect of the $\alpha$ parameter on the clustering signals and that of the scatter in abundance matching are degenerate, here we present the joint constraints on $\alpha$ and scatter using the SDSS galaxy catalog.

We compute the $\chi^2$ statistic to evaluate the goodness-of-fit for a set of values in the $(\alpha, \text{scatter})$ parameter space for each threshold. We also compute the $\chi^2$ statistic for several different cosmological boxes to determine whether the constraint on $(\alpha, \text{scatter})$ varies significantly between boxes using different cosmologies and codes.

The $\chi^2$ statistic is computed as
\begin{equation}
\chi^2 = \sum_i \sum_j d(r_p^i) d(r_p^j) C^{-1}(r_p^i, r_p^j),
\end{equation}
where $d(r_p) := w_p^\text{mock}(r_p) - w_p^\text{SDSS}(r_p)$, $C(r_p^i, r_p^j)$ is the covariance, and $i, j$ denote the indices of the bins of $r_p$.
Note that here $C$ already includes the covariance from the SDSS data, as well as the covariance from jackknifing the mock catalog and from multiple realizations of abundance matching.

\added{Note that in our analysis, we compute the $p$-value using the $\chi^2$ test independently for each set of parameter values. Hence the $p$-values presented in the figures below are not comparisons between different parameter values, but always are with respect to the null hypothesis.} 

\begin{figure*}[htb!]
\putfig{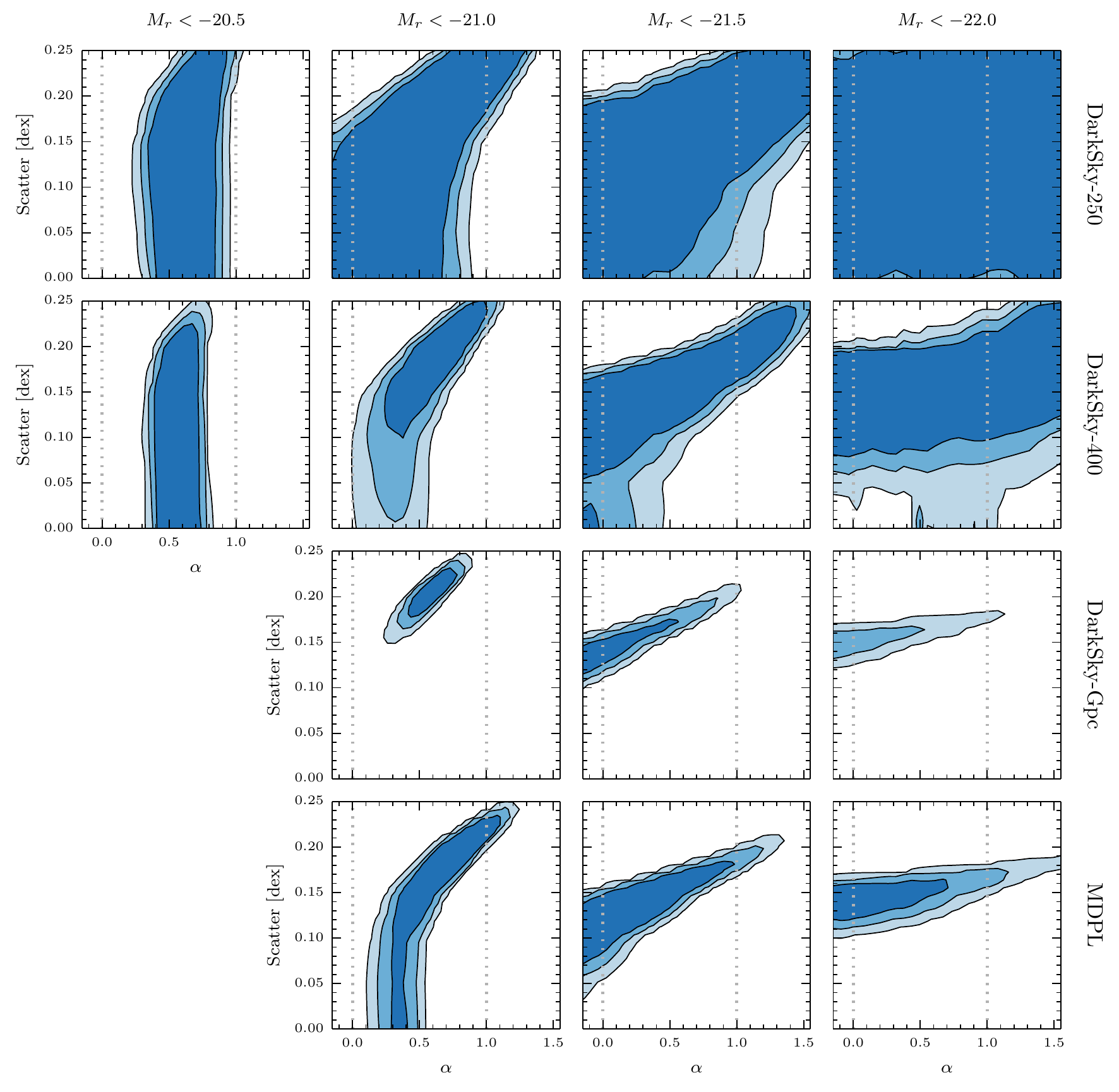}
\caption{Constraints on $\alpha$ and scatter in each of the three \texttt{DarkSky} boxes and the \texttt{MDPL} box, for four thresholds ($M_r < -20.5$, $-21$, $-21.5$, and $-22$; from left to right). The contours, from dark to light blue, show the one-side $p$-value of 0.05, 0.01, and 0.001 for the $\chi^2$ fit.}
\label{fig:darksky-boxes}
\end{figure*}

\autoref{fig:darksky-boxes} shows the constraints from the three \texttt{DarkSky} boxes and the \texttt{MDPL} box for four different luminosity thresholds separately. 
Note that the two $1\Gpch{}$ boxes do not have the resolution to generate a complete sample below roughly $M_r = -21$, and hence we omit the lowest luminosity panels for these boxes in \autoref{fig:darksky-boxes}. 
We will discuss detailed resolution requirements for abundance matching in upcoming work.

We see several interesting features here. First, the degeneracy between $\alpha$ and scatter is most visible in the samples of $M_r < -21.5$ and $-21$. In both cases, we see the degeneracy as expected: a larger $\alpha$ requires a larger scatter to balance the additional clustering since more highly concentrated halos are included.

Second, as expected, larger boxes provide stronger constraints, indicating that the constraining power of most previous studies, which have almost exclusively used boxes of $\sim 250\Mpch$ on a side, have been dominated by sample variance. This is especially true for the brighter samples because the numbers of galaxies in those samples are small.  While the sample of $M_r < -22$ from \texttt{DarkSky-250} and \texttt{-400} provide little constraint on $\alpha$ and scatter, the samples from the $1\Gpch{}$ boxes give clear constraints on scatter, and exclude zero scatter in this range of $\alpha$ at $p < 0.001$.

Third, on the faint end, we obtain a much stronger constraint on $\alpha$.  The luminosity dependence of halo bias is significantly weaker in this regime, and thus these galaxies do not provide strong constraints on scatter.  However, with \texttt{DarkSky-400}, this sample excludes both $\alpha=0$ and $1$ at $p < 0.001$.

\subsection{Combining Constraints from Different Thresholds}
\label{sec:am-alpha-scatter}

\begin{figure}[htb!]
\putfig{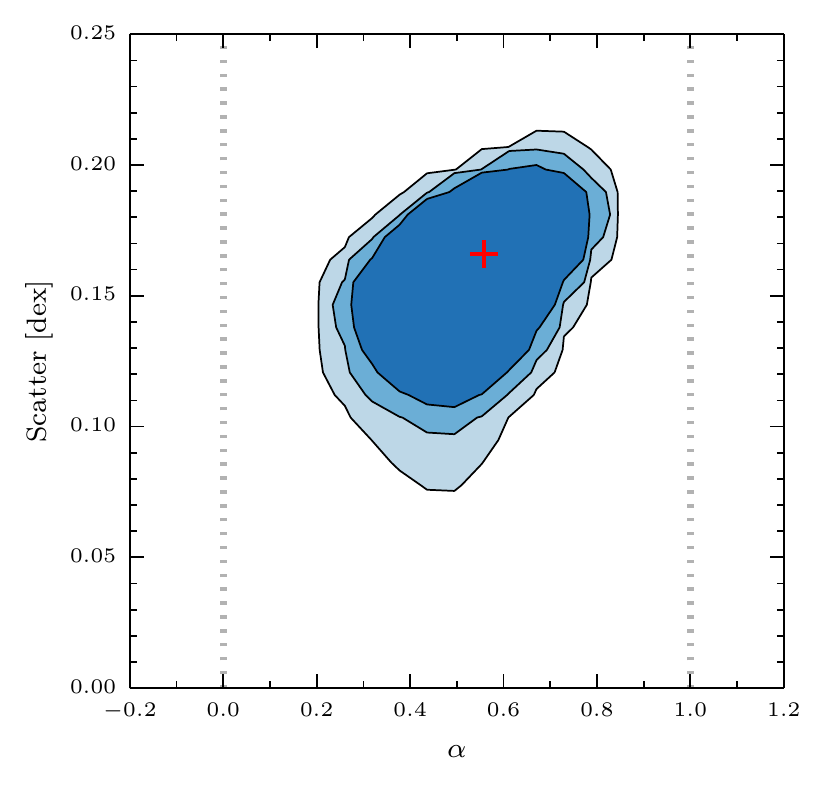}
\caption{Combined joint constraint on $\alpha$ and scatter 
from four thresholds ($M_r < -20.5$, $-21$, $-21.5$, and $-22.0$) for \texttt{DarkSky-400}. The contours, from dark to light blue, show the one-side $p$-value of 0.05, 0.01, and 0.001 for the $\chi^2$ fit.  Crosshairs show best-fit point ($\alpha = 0.57^{+0.20}_{-0.27}$; scatter $= 0.17^{+0.03}_{-0.05}$ dex).}
\label{fig:combined-constraint-ds14-i}
\end{figure}

\begin{figure*}[htb!]
\putfig{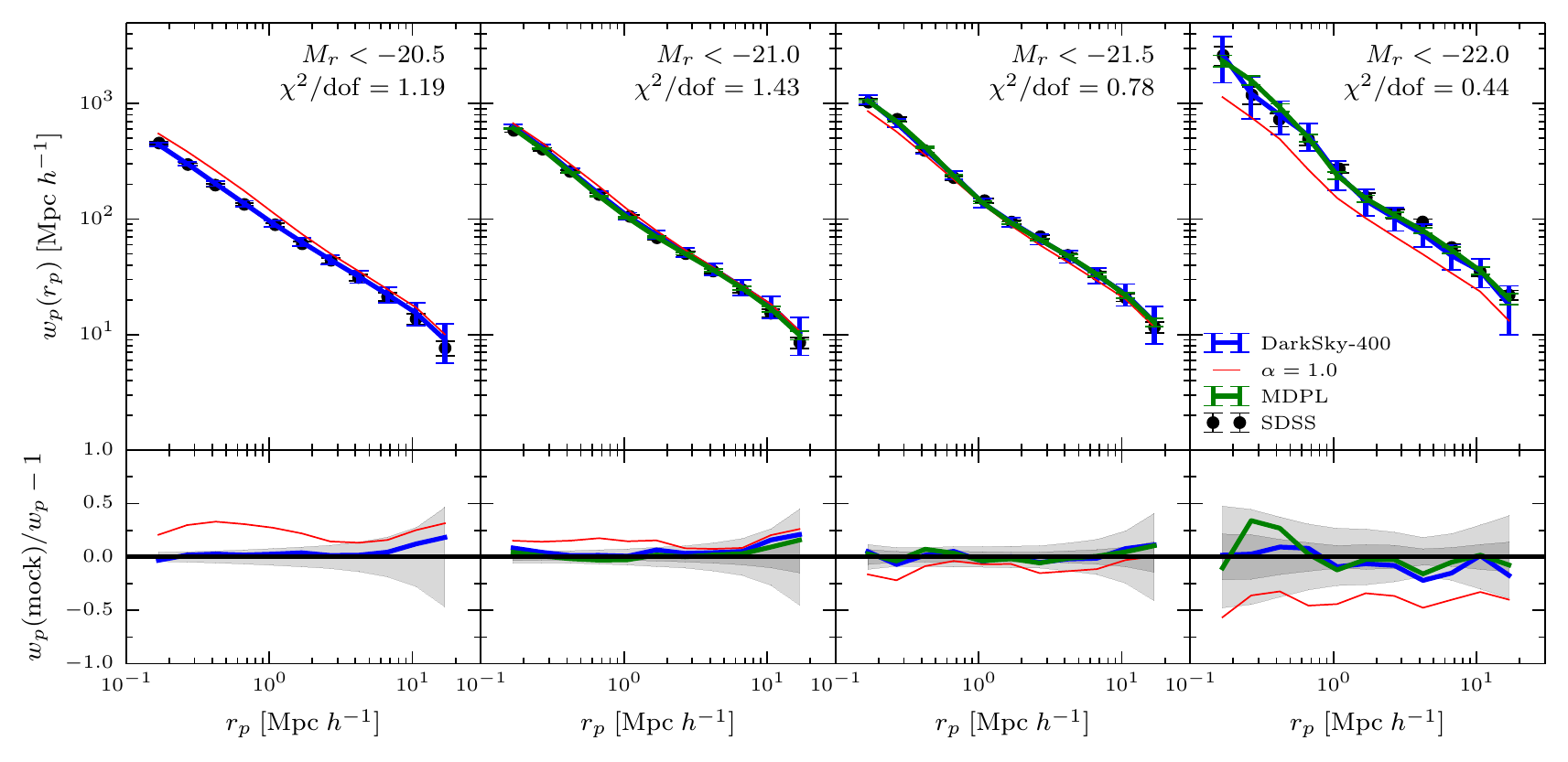}
\caption{Comparison between our best-fit model and SDSS data.
Top row shows the best-fit $\wprp$ for \texttt{DarkSky-400} (blue line; $\alpha = 0.57$; scatter $=$ 0.17 dex) and \texttt{MDPL} (green line; $\alpha = 0.49$; scatter $=$ 0.16 dex). The $\chi^2$ values are shown for \texttt{DarkSky-400} at each threshold. Circles show SDSS $\wprp$.
Four columns represent four thresholds ($M_r < -20.5$, $-21$, $-21.5$, and $-22$; from left to right).
To compare with previous work, the red line shows the $\hat v_{1.0}$ model with best-fit scatter of 0.22 dex in \texttt{DarkSky-400} ($\chi^2/\text{dof}=$ 8.9, 2.5, 1.9, and 1.8 for the four thresholds respectively).
\texttt{MDPL} is omitted from the $M_r<-20.5$ column, which is not used in fit computation for that box.
Bottom row shows the relative difference with respect to SDSS data.
Gray bands indicate combined SDSS and mock errors: the outer region indicates combined errors for \texttt{DarkSky-400}, while the inner region indicates combined errors for \texttt{MDPL}.}
\label{fig:single-best-fit-ds14i-NewMDPL}
\end{figure*}

If we assume that $\alpha$ and scatter are constant with respect to luminosity, then the samples at different thresholds can be combined to produce an overall constraint on $\alpha$ and scatter.  Here, we also assume that the constraints from samples at different thresholds are independent. This assumption is only approximately correct for two reasons. First, although the sample at each threshold is dominated by the fainter galaxies, it does include galaxies from higher thresholds. Second, for a given simulation, the clustering signals at different thresholds are also correlated. Here we assume the independence for simplicity, and because the effects are both small, we do not expect that they will significantly impact our conclusions.

The combined constraint from four thresholds ($M_r < -20.5$, $-21$, $-21.5$, and $-22$) for \texttt{DarkSky-400} is shown in \autoref{fig:combined-constraint-ds14-i}. 
This combined constraint excludes both $\alpha=0$ (resembling $\sub{M}{peak}$) and $1$ (resembling $\sub{v}{peak}$) at $p < 0.001$, and also excludes zero scatter at $p < 0.001$.
The best-fit values for \texttt{DarkSky-400} are $\alpha = 0.57^{+0.20}_{-0.27}$ and scatter $=0.17^{+0.03}_{-0.05}$ dex.
This value of $\alpha$ is consistent with the value that best matches the observed satellite fraction shown in \autoref{sec:discussion-fsat}.

The $\wprp$ corresponding to this best-fit model is shown in \autoref{fig:single-best-fit-ds14i-NewMDPL}. We find that, with this new $\hat{v}_\alpha$ proxy, we can reproduce the $\wprp$ observed in the SDSS data at all four luminosity thresholds very closely, with a fixed value of $\alpha$ and scatter. In the same figure, the best-fit $\wprp$ for \texttt{MDPL} is also shown. The large size of the \texttt{MDPL} box results in much smaller errors on the mock $\wprp$, yet we still obtain excellent agreement with observations.  We note that the agreement is good down to the small scales measured by \cite{Zehavi2011}.

\subsection{Consistency between Different Simulations}

\begin{figure}[htb!]
\putfig{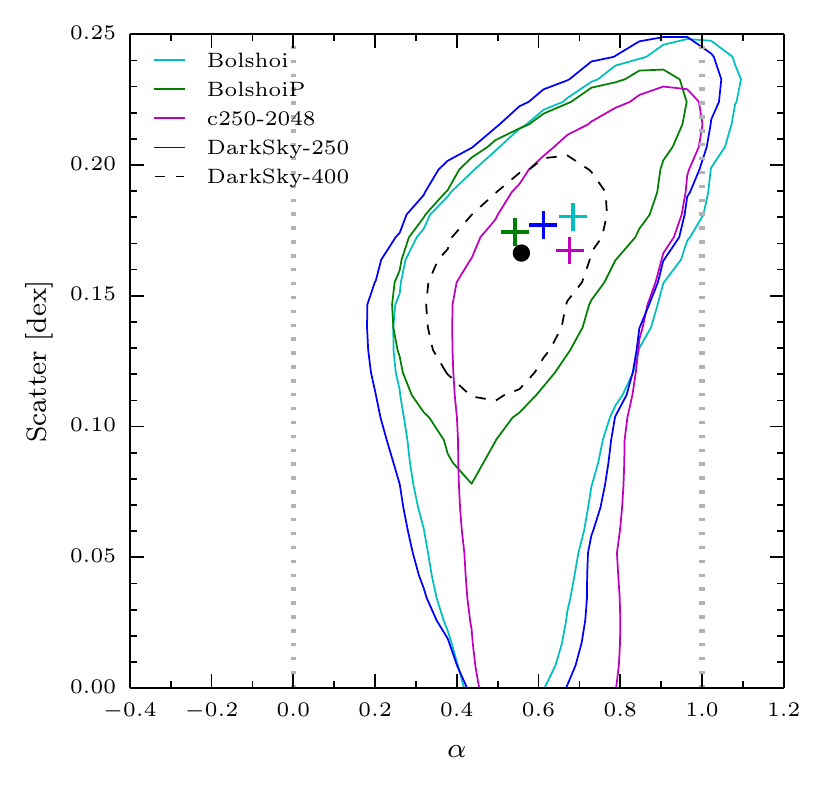}
\caption{Contours of $p$-value = 0.05 for the combined samples ($M_r < -20.5$, $-21$, and $-21.5$), for different simulations: \texttt{DarkSky-250} (blue), \texttt{c250-2048} (magenta), \texttt{Bolshoi} (cyan), \texttt{BolshoiP} (green), and \texttt{DarkSky-400} (black dashed). Crosshairs and the dot show best-fit points for the corresponding boxes.}
\label{fig:1-sigma-contours-250}
\end{figure}

We repeat our study on the clustering with the other simulations listed in \autoref{tab:cosmosim} to test the robustness of our results and to investigate their cosmology dependence. We use four $250\Mpch{}$ boxes with approximately the same mass resolution but with different cosmologies, three different $N$-body codes, and different initial conditions.

\autoref{fig:1-sigma-contours-250} shows the $p > 0.05$ regions in $(\alpha, \text{scatter})$ from these four boxes, and also \texttt{DarkSky-400} for reference.
Despite the difference between these boxes, the $p = 0.05$ contours agree with one another very well, and the best-fit points are all in proximity in this parameter space. This result demonstrates the robustness of our analysis. It also suggests that, within the range of cosmologies tested here (all modern $\Lambda$CDM cosmologies but with a range of values of, e.g., $\Omega_M$ and $\sigma_8$), the cosmology dependence is weak enough that it cannot be distinguished in these $250\Mpch{}$ boxes.

\subsection{Application to Dimmer Galaxy Samples}
\label{sec:dimmer}

\begin{figure*}[htb!]
  \putfig{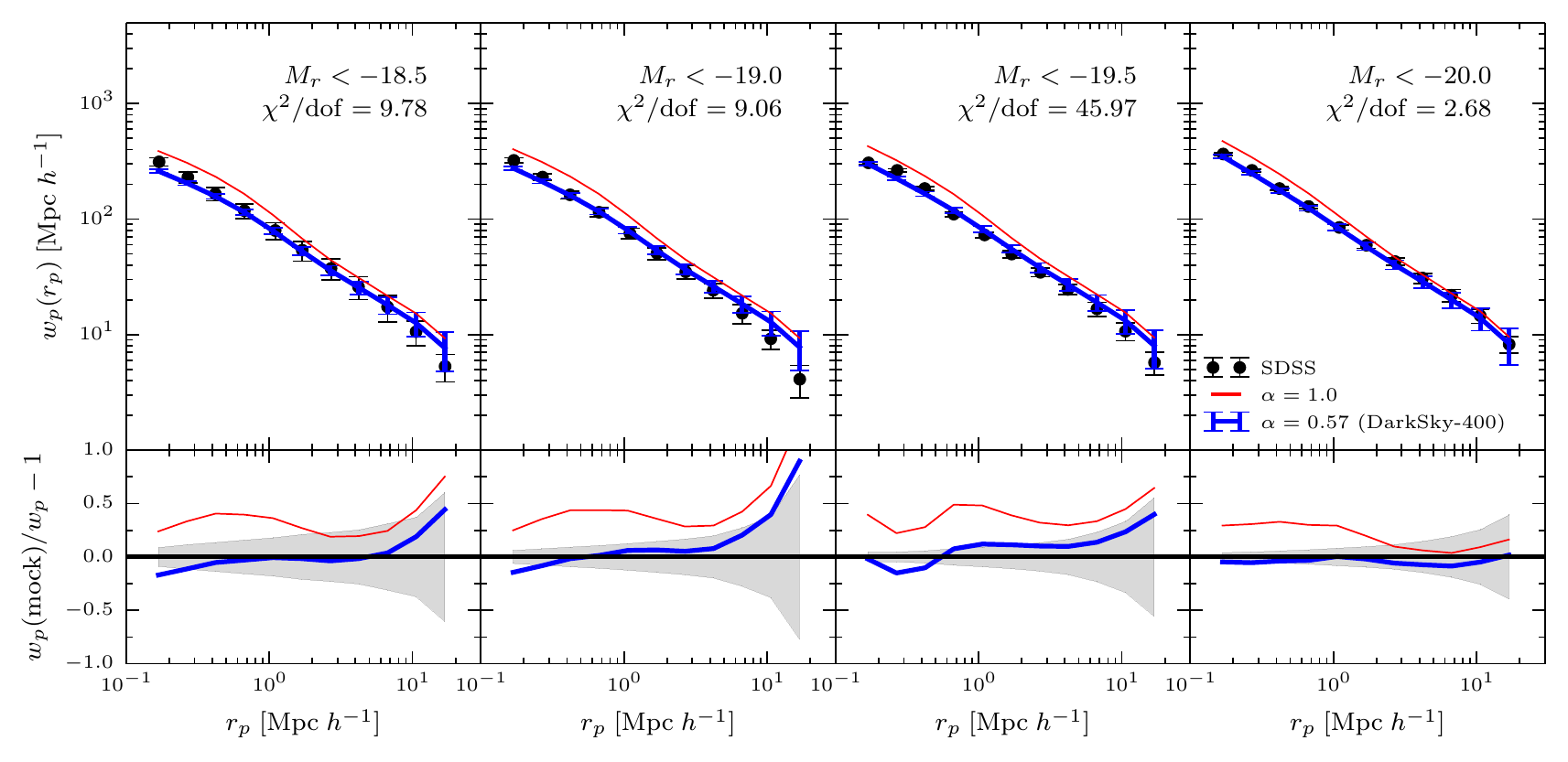}
  \caption{Comparison between our best-fit model and SDSS data, for dimmer thresholds than are used in the fit.
Top row shows $\wprp$ for \texttt{DarkSky-400} with $\alpha = 0.57$ and scatter $=$ 0.17 dex. The $\chi^2$ values are also shown for this model at each threshold. These values are the same as those used in \autoref{fig:single-best-fit-ds14i-NewMDPL}. Circles show SDSS $\wprp$.
   Four columns represent four dimmer thresholds ($M_r < -18.5$, $-19$, $-19.5$, and $-20$; from left to right).
   To compare with previous work, the red line shows $\hat v_{1.0}$ model with a scatter of 0.22 dex in \texttt{DarkSky-400}.
   Bottom row shows the relative difference with respect to SDSS data.
   Gray bands indicate combined SDSS and mock errors.}
  \label{fig:dimmer-best-fit-ds14i}
\end{figure*}

Since abundance matching models are based on the halo catalogs of $N$-body simulations, they suffer from the same limitations due to finite resolution.
Particularly, for dimmer samples, abundance matching models tend to underpredict small-scale clustering \citep{Guo2015}.
To avoid possible impact of the limited resolution and to obtain unbiased constraints on $\alpha$ and scatter, we only used galaxy samples brighter than $M_r = -20.5$ in the main results.
We demonstrate that $\alpha$ can already be constrained even with only the bright samples. 

Nevertheless, the model presented here can also provide good fits to dimmer galaxy samples given its flexibility.
Even with the best-fit values of $\alpha$ and scatter used in \autoref{fig:single-best-fit-ds14i-NewMDPL} (i.e., solely from the bright samples), we can obtain reasonably good matches to the clustering of dimmer samples, as shown in \autoref{fig:dimmer-best-fit-ds14i}.
We note that for these four dimmer samples with $M_r > -20.5$, observed galaxies are more clustered at small scales and less so at large scales when compared to  the model prediction with these particular parameter values. 
This hints at larger values of both $\alpha$ and scatter, and hence at the mass dependence of $\alpha$ and scatter. 
However, this hint could be a result of the bias due to resolution limit, and simulations of higher resolution are needed to obtain a definitive conclusion on this possible mass dependence of $\alpha$ and scatter. 

\section{Discussion}
\label{sec:discussion}
\subsection{Consistency with Previous Work}

We note that \cite{Reddick2013} are able to get reasonable fits to the clustering measurements by abundance matching to $\sub{v}{peak}$.  However, the amount of scatter required in the $\sub{v}{peak}$ case is large (0.22 dex) compared to other constraints in the literature \citep[e.g.,][]{More2009}. Additionally, this $\sub{v}{peak}$ model requires the exclusion of subhalos whose current mass is less than some fraction of the peak mass (using the parameter $\sub{\mu}{cut} := \sub{M}{vir,now}/\sub{M}{peak}$), and we do not find this to be required with $\hat{v}_\alpha$.
Furthermore, the $\sub{v}{peak}$ model did not provide a good fit to the brightest samples when matching to luminosity \citep[see the top left panel of Figure 26 of][]{Reddick2013}, nor did it fit the satellite fraction without excluding halos of low $\sub{M}{vir,now}/\sub{M}{peak}$ (i.e., with $\sub{\mu}{cut}=0$ in \citealt[Figure 22]{Reddick2013}).

The analysis in our present paper uses a larger box with about four times the volume, and thus provides more constraining power. In \autoref{fig:1-sigma-contours-250}, one can see that given the degeneracy between scatter and $\alpha$, the smaller Bolshoi box does allow for a region with $\alpha = 1$ (corresponding to $\sub{v}{peak}$) with higher scatter ($\gtrsim$~0.2).
This region is consistent with the best-fit result of \citet{Reddick2013}, but is ruled out here with the larger \texttt{DarkSky-400} box.

\subsection{Flexibility of the Abundance Matching Framework}

The core idea of abundance matching is two key assumptions: (1) all galaxies live in dark matter density peaks, and (2) galaxy properties are well correlated with halo properties.  However, abundance matching should not be viewed as a ``parameter-free'' model, but instead, 
can be viewed as a flexible description of the galaxy--halo connection whose parameters can be constrained by observations. 

By introducing this new interpolation scheme with the parameter $\alpha$, we demonstrate that the abundance matching technique is more flexible than the version that was originally proposed. This interpolation scheme also provides a novel interpretation of the matching proxy.
Traditionally, when we compare the performance of two abundance matching proxies, we tend to overemphasize the physical meaning of the proxy that performs better. With this $\alpha$ parameter, we demonstrate that, under the framework of abundance matching, there is indeed nothing special about the maximal circular velocity. It is only that observations of clustering statistics favor more concentration dependence than using simply halo mass as a proxy.

On a different note, the $\alpha$ parameter affects the galaxy clustering in the resulting catalog by changing the satellite fraction and the amount of assembly bias.
However, we also note that, within the framework of abundance matching, these two effects (assembly bias and satellite fraction) are \emph{linked} in the specific way when one changes the parameter $\alpha$. This link is physically justified if all galaxies live in resolved halos and if galaxy and halo  properties can be effectively rank matched with one of the proxies considered.
On the contrary, the model in \citet{Hearin2016} do \emph{not} assume this link, and the two effects can be adjusted separately. 
Nevertheless, with the clustering statistics we tested here, there is no evidence that this link, implicitly assumed when one uses abundance matching, needs to be broken.

This linked feature also enables us to constrain $\alpha$ with only the two-halo clustering. In fact, when we exclude small scales in our analysis, we obtain a consistent, though weaker, constraint on $\alpha$. This is promising due to the more difficult nature of modeling the smallest scales, which can be impacted by fiber collisions in the data, and by resolution and baryonic effects in the simulations. 
At present, our best constraint on $\alpha$ still comes from scales in the one-halo regime, but stronger large scale constraints will be possible as data samples become larger. This result suggests that many of the key details of the galaxy--halo connection may be constrained even without the smallest scales.

It is also important to note that, in addition to the concentration dependence discussed in this work, there is still a rich set of parameters that can potentially be included in abundance matching without breaking the core assumptions mentioned above, such as using non-constant or non-Gaussian scatter, evaluating the matching proxy at different epochs, and adopting different treatments for central and satellite galaxies. 
With future simulations that have larger volumes and higher resolutions, we can constrain these potential abundance matching parameters, and in return obtain insights on the physical processes of galaxy formation.

\subsection{Constraining Power from Other Statistics}

Several other probes can provide complementary constraining power on the $\alpha$ parameter. Although in this paper we have not completed a full analysis of satellite fractions, \autoref{fig:fsat-vs-alpha} already demonstrates that the satellite fraction as a function of luminosity can provide independent constraints on $\alpha$. 
Similarly, other group statistics, such as the conditional luminosity function, should also provide additional constraints on $\alpha$ and scatter.  

As an example, R.~M.~Reddick et al.\ (2016, in preparation) have studied the conditional luminosity function of galaxies in the redMaPPer cluster sample \citep{Rykoff2014}. This sample consists of a very large number of photometrically identified clusters, and hence allows for very small statistical errors on the parameters. This work finds that for models with lower scatter, data require a stronger anti-correlation between satellite occupation and central luminosity. Since satellite occupation is also anti-correlated with host halo concentration \citep{Zentner2005,Mao2015}, the result of R.~M.~Reddick et al.\ (2016, in preparation) implies an anti-correlation between scatter and $\alpha$ (i.e., the concentration dependence of luminosity). This result would then be complementary to the clustering results presented here, since the latter finds a positive correlation between scatter and $\alpha$, provided that the correlation between scatter and $\alpha$ behaves the same in both luminosity-selected and redMaPPer samples.

Although we do not investigate this directly here, measurements of galaxy voids may be able to put further constraints on the amount of assembly bias  \citep{Tinker2008, Tinker2009}.  Combining clustering results with measurements of galaxy--galaxy lensing may be able to put further limits on the scatter and on the concentration dependence \citep[e.g.,][]{Tasitsiomi2004, Mandelbaum2006, Neistein2012}. In addition, data that have more information on redshift dependence, such as the pseudo-multipole correlation function, can provide more constraints on these parameters \citep{Reid2014,Guo2015,Saito2015}.

Another way to put a physical prior on the parameters in empirical models is to compare the model predictions with hydrodynamic simulations. For example, \cite{Chaves-Montero2015} evaluated the galaxy--halo connection in the EAGLE simulation with various abundance matching models with different epochs at which the matching proxy is evaluated. In this work, we establish that the galaxy luminosity has at least some dependence on halo concentration.  It will be interesting to fully understand whether and to what extent such a luminosity dependence arises in modern hydrodynamic simulations, and what physical parameters it depends on.

\section{Summary} \label{sec:summary} 
We introduce a generalization of abundance matching that allows adjustable concentration dependence. In particular, we propose a model that abundance matches to a parameter $\hat{v}_\alpha$, which smoothly interpolates between  $v_{\text{vir}}$ (when $\alpha=0$) and $v_{\text{max}}$ (when $\alpha=1$), both of which are evaluated at the peak value of the mass accretion histories. 

Within the framework of abundance matching, the parameter $\alpha$ controls the concentration dependence of luminosity at given halo mass. Hence, $\alpha$ impacts both the satellite fraction and the assembly bias in the resulting mock galaxy catalog.  Both effects lead to larger clustering for higher values of $\alpha$, but the satellite fraction primarily increases clustering at small scales (the one-halo term), while assembly bias primarily increases clustering at larger scales (the two-halo term).
This model is the first to introduce a continuously adjustable assembly bias within the abundance matching framework.

We further demonstrate that the current clustering measurements from SDSS already have constraining power on this parameter $\alpha$. 
SDSS data prefer a range of $\alpha$ in the region between 0 and 1, i.e., with $\hat{v}_\alpha$ between $v_{\text{vir, peak}}$ and $v_{\text{max, peak}}$. Our best-fit value is $ \alpha = 0.57^{+0.20}_{-0.27}$, with a scatter value of $0.17^{+0.03}_{-0.05}$ dex.
With the high-resolution 400$\Mpch$ box, \texttt{DarkSky-400}, we show that the halo parameters $M_{\text{vir, peak}}$ and $v_{\text{max, peak}}$, which have been previously used in the literature are both ruled out at $p < 0.001$ when the various luminosity thresholds are combined.

In conclusion, the more general abundance matching model we present here is an important step in the quest for precise and accurate models of galaxy clustering down to small scales, which will be essential to take full advantage of the next generation of cosmological surveys.

\acknowledgments{B.V.L.\ was supported by a Stanford University undergraduate research grant. Y.Y.M.\ was supported by the Weiland Family Stanford Graduate Fellowship. This work received partial support from the U.S.\ Department of Energy under contract number DE-AC02-76SF00515.  We thank Harry Desmond and Andrew Hearin for useful discussions.
We thank Rachel Reddick for providing the measurement of observed galaxy statistics used in this work.  This research made use of the Dark Sky Simulations, which were produced using an INCITE 2014 allocation on the Oak Ridge Leadership Computing Facility at Oak Ridge National Laboratory.  We thank Mike Warren, Matt Turk, and our other Dark Sky collaborators for their efforts in creating these simulations and for providing access to them. We thank Peter Behroozi for providing the halo catalogs and merger trees for the \texttt{Bolshoi}, \texttt{BolshoiP}, and \texttt{MDPL} simulations, which were run by the MultiDark Collaboration.  Additional simulations and computations were performed using computational resources at SLAC and at NERSC. We thank the SLAC computational team for their consistent support.}

\bibliographystyle{yahapj}
\bibliography{bibliography}

\begin{thebibliography}{}
\providecommand\natexlab[1]{#1}
\providecommand\JournalTitle[1]{#1}

\bibitem[{{Abazajian} {et~al.}(2009){Abazajian}, {Adelman-McCarthy},
  {Ag{\"u}eros}, {Allam}, {Allende Prieto}, {An}, {Anderson}, {Anderson},
  {Annis}, {Bahcall}, \& et~al.}]{Abazajian2009}
{Abazajian}, K.~N., {Adelman-McCarthy}, J.~K., {Ag{\"u}eros}, M.~A., {et~al.}
  2009,
  \href{http://dx.doi.org/10.1088/0067-0049/182/2/543}{\JournalTitle{\apjs},
  182, 543}

\bibitem[{{Behroozi} {et~al.}(2010){Behroozi}, {Conroy}, \&
  {Wechsler}}]{Behroozi2010}
{Behroozi}, P.~S., {Conroy}, C., \& {Wechsler}, R.~H. 2010,
  \href{http://dx.doi.org/10.1088/0004-637X/717/1/379}{\JournalTitle{\apj},
  717, 379}

\bibitem[{{Behroozi} {et~al.}(2014){Behroozi}, {Wechsler}, {Lu}, {Hahn},
  {Busha}, {Klypin}, \& {Primack}}]{Behroozi2014}
{Behroozi}, P.~S., {Wechsler}, R.~H., {Lu}, Y., {et~al.} 2014,
  \href{http://dx.doi.org/10.1088/0004-637X/787/2/156}{\JournalTitle{\apj},
  787, 156}

\bibitem[{{Behroozi} {et~al.}(2013{\natexlab{a}}){Behroozi}, {Wechsler}, \&
  {Wu}}]{Behroozi2013a}
{Behroozi}, P.~S., {Wechsler}, R.~H., \& {Wu}, H.-Y. 2013{\natexlab{a}},
  \href{http://dx.doi.org/10.1088/0004-637X/762/2/109}{\JournalTitle{\apj},
  762, 109}

\bibitem[{{Behroozi} {et~al.}(2013{\natexlab{b}}){Behroozi}, {Wechsler}, {Wu},
  {Busha}, {Klypin}, \& {Primack}}]{Behroozi2013b}
{Behroozi}, P.~S., {Wechsler}, R.~H., {Wu}, H.-Y., {et~al.} 2013{\natexlab{b}},
  \href{http://dx.doi.org/10.1088/0004-637X/763/1/18}{\JournalTitle{\apj}, 763,
  18}

\bibitem[{{Berlind} \& {Weinberg}(2002)}]{Berlind2002}
{Berlind}, A.~A., \& {Weinberg}, D.~H. 2002,
  \href{http://dx.doi.org/10.1086/341469}{\JournalTitle{\apj}, 575, 587}

\bibitem[{{Blanton} {et~al.}(2005){Blanton}, {Schlegel}, {Strauss},
  {Brinkmann}, {Finkbeiner}, {Fukugita}, {Gunn}, {Hogg}, {Ivezi{\'c}}, {Knapp},
  {Lupton}, {Munn}, {Schneider}, {Tegmark}, \& {Zehavi}}]{Blanton2005}
{Blanton}, M.~R., {Schlegel}, D.~J., {Strauss}, M.~A., {et~al.} 2005,
  \href{http://dx.doi.org/10.1086/429803}{\JournalTitle{\aj}, 129, 2562}

\bibitem[{{Bower} {et~al.}(2006){Bower}, {Benson}, {Malbon}, {Helly}, {Frenk},
  {Baugh}, {Cole}, \& {Lacey}}]{Bower2006}
{Bower}, R.~G., {Benson}, A.~J., {Malbon}, R., {et~al.} 2006,
  \href{http://dx.doi.org/10.1111/j.1365-2966.2006.10519.x}{\JournalTitle{\mnras},
  370, 645}

\bibitem[{{Bryan} \& {Norman}(1998)}]{Bryan1998}
{Bryan}, G.~L., \& {Norman}, M.~L. 1998,
  \href{http://dx.doi.org/10.1086/305262}{\JournalTitle{\apj}, 495, 80}

\bibitem[{{Bullock} {et~al.}(2002){Bullock}, {Wechsler}, \&
  {Somerville}}]{Bullock2002}
{Bullock}, J.~S., {Wechsler}, R.~H., \& {Somerville}, R.~S. 2002,
  \href{http://dx.doi.org/10.1046/j.1365-8711.2002.04959.x}{\JournalTitle{\mnras},
  329, 246}

\bibitem[{{Chaves-Montero} {et~al.}(2016){Chaves-Montero}, {Angulo}, {Schaye},
  {Schaller}, {Crain}, {Furlong}, \& {Theuns}}]{Chaves-Montero2015}
{Chaves-Montero}, J., {Angulo}, R.~E., {Schaye}, J., {et~al.} 2016,
  \href{http://dx.doi.org/10.1093/mnras/stw1225}{\JournalTitle{\mnras}, 460,
  3100}

\bibitem[{{Conroy} {et~al.}(2006){Conroy}, {Wechsler}, \&
  {Kravtsov}}]{Conroy2006}
{Conroy}, C., {Wechsler}, R.~H., \& {Kravtsov}, A.~V. 2006,
  \href{http://dx.doi.org/10.1086/503602}{\JournalTitle{\apj}, 647, 201}

\bibitem[{{Cooray} \& {Sheth}(2002)}]{Cooray2002}
{Cooray}, A., \& {Sheth}, R. 2002,
  \href{http://dx.doi.org/10.1016/S0370-1573(02)00276-4}{\JournalTitle{\physrep},
  372, 1}

\bibitem[{{Crain} {et~al.}(2015){Crain}, {Schaye}, {Bower}, {Furlong},
  {Schaller}, {Theuns}, {Dalla Vecchia}, {Frenk}, {McCarthy}, {Helly},
  {Jenkins}, {Rosas-Guevara}, {White}, \& {Trayford}}]{2015MNRAS.450.1937C}
{Crain}, R.~A., {Schaye}, J., {Bower}, R.~G., {et~al.} 2015,
  \href{http://dx.doi.org/10.1093/mnras/stv725}{\JournalTitle{\mnras}, 450,
  1937}

\bibitem[{{Croton} {et~al.}(2007){Croton}, {Gao}, \& {White}}]{Croton2007}
{Croton}, D.~J., {Gao}, L., \& {White}, S.~D.~M. 2007,
  \href{http://dx.doi.org/10.1111/j.1365-2966.2006.11230.x}{\JournalTitle{\mnras},
  374, 1303}

\bibitem[{{Croton} {et~al.}(2006){Croton}, {Springel}, {White}, {De Lucia},
  {Frenk}, {Gao}, {Jenkins}, {Kauffmann}, {Navarro}, \& {Yoshida}}]{Croton2006}
{Croton}, D.~J., {Springel}, V., {White}, S.~D.~M., {et~al.} 2006,
  \href{http://dx.doi.org/10.1111/j.1365-2966.2005.09675.x}{\JournalTitle{\mnras},
  365, 11}

\bibitem[{{Desmond} \& {Wechsler}(2015)}]{Desmond2015}
{Desmond}, H., \& {Wechsler}, R.~H. 2015,
  \href{http://dx.doi.org/10.1093/mnras/stv1978}{\JournalTitle{\mnras}, 454,
  322}

\bibitem[{{Gao} {et~al.}(2005){Gao}, {White}, {Jenkins}, {Frenk}, \&
  {Springel}}]{Gao2005}
{Gao}, L., {White}, S.~D.~M., {Jenkins}, A., {Frenk}, C.~S., \& {Springel}, V.
  2005,
  \href{http://dx.doi.org/10.1111/j.1365-2966.2005.09509.x}{\JournalTitle{\mnras},
  363, 379}

\bibitem[{{Guo} {et~al.}(2016{\natexlab{a}}){Guo}, {Zheng}, {Behroozi},
  {Zehavi}, {Comparat}, {Favole}, {Gottl{\"o}ber}, {Klypin}, {Prada},
  {Rodr{\'{\i}}guez-Torres}, {Weinberg}, \& {Yepes}}]{1608.03660}
{Guo}, H., {Zheng}, Z., {Behroozi}, P.~S., {et~al.} 2016{\natexlab{a}},
  \href{http://dx.doi.org/10.3847/0004-637X/831/1/3}{\JournalTitle{\apj}, 831,
  3}

\bibitem[{{Guo} {et~al.}(2016{\natexlab{b}}){Guo}, {Zheng}, {Behroozi},
  {Zehavi}, {Chuang}, {Comparat}, {Favole}, {Gottloeber}, {Klypin}, {Prada},
  {Rodr{\'{\i}}guez-Torres}, {Weinberg}, \& {Yepes}}]{Guo2015}
---. 2016{\natexlab{b}},
  \href{http://dx.doi.org/10.1093/mnras/stw845}{\JournalTitle{\mnras}, 459,
  3040}

\bibitem[{{Hearin} {et~al.}(2014){Hearin}, {Watson}, {Becker}, {Reyes},
  {Berlind}, \& {Zentner}}]{Hearin2014}
{Hearin}, A.~P., {Watson}, D.~F., {Becker}, M.~R., {et~al.} 2014,
  \href{http://dx.doi.org/10.1093/mnras/stu1443}{\JournalTitle{\mnras}, 444,
  729}

\bibitem[{{Hearin} {et~al.}(2013){Hearin}, {Zentner}, {Berlind}, \&
  {Newman}}]{Hearin2013a}
{Hearin}, A.~P., {Zentner}, A.~R., {Berlind}, A.~A., \& {Newman}, J.~A. 2013,
  \href{http://dx.doi.org/10.1093/mnras/stt755}{\JournalTitle{\mnras}, 433,
  659}

\bibitem[{{Hearin} {et~al.}(2016){Hearin}, {Zentner}, {van den Bosch},
  {Campbell}, \& {Tollerud}}]{Hearin2016}
{Hearin}, A.~P., {Zentner}, A.~R., {van den Bosch}, F.~C., {Campbell}, D., \&
  {Tollerud}, E. 2016,
  \href{http://dx.doi.org/10.1093/mnras/stw840}{\JournalTitle{\mnras}, 460,
  2552}

\bibitem[{{Henriques} {et~al.}(2015){Henriques}, {White}, {Thomas}, {Angulo},
  {Guo}, {Lemson}, {Springel}, \& {Overzier}}]{Henriques2015}
{Henriques}, B.~M.~B., {White}, S.~D.~M., {Thomas}, P.~A., {et~al.} 2015,
  \href{http://dx.doi.org/10.1093/mnras/stv705}{\JournalTitle{\mnras}, 451,
  2663}

\bibitem[{{Klypin} {et~al.}(2016){Klypin}, {Yepes}, {Gottl{\"o}ber}, {Prada},
  \& {He{\ss}}}]{Klypin2014}
{Klypin}, A., {Yepes}, G., {Gottl{\"o}ber}, S., {Prada}, F., \& {He{\ss}}, S.
  2016, \href{http://dx.doi.org/10.1093/mnras/stw248}{\JournalTitle{\mnras},
  457, 4340}

\bibitem[{{Klypin} {et~al.}(2011){Klypin}, {Trujillo-Gomez}, \&
  {Primack}}]{Klypin2011}
{Klypin}, A.~A., {Trujillo-Gomez}, S., \& {Primack}, J. 2011,
  \href{http://dx.doi.org/10.1088/0004-637X/740/2/102}{\JournalTitle{\apj},
  740, 102}

\bibitem[{{Kravtsov} {et~al.}(2004){Kravtsov}, {Berlind}, {Wechsler}, {Klypin},
  {Gottl{\"o}ber}, {Allgood}, \& {Primack}}]{Kravtsov2004}
{Kravtsov}, A.~V., {Berlind}, A.~A., {Wechsler}, R.~H., {et~al.} 2004,
  \href{http://dx.doi.org/10.1086/420959}{\JournalTitle{\apj}, 609, 35}

\bibitem[{{Kulier} \& {Ostriker}(2015)}]{Kulier2015}
{Kulier}, A., \& {Ostriker}, J.~P. 2015,
  \href{http://dx.doi.org/10.1093/mnras/stv1564}{\JournalTitle{\mnras}, 452,
  4013}

\bibitem[{{Lin} {et~al.}(2016){Lin}, {Mandelbaum}, {Huang}, {Huang}, {Dalal},
  {Diemer}, {Jian}, \& {Kravtsov}}]{Lin2015}
{Lin}, Y.-T., {Mandelbaum}, R., {Huang}, Y.-H., {et~al.} 2016,
  \href{http://dx.doi.org/10.3847/0004-637X/819/2/119}{\JournalTitle{\apj},
  819, 119}

\bibitem[{{Lu} {et~al.}(2014){Lu}, {Wechsler}, {Somerville}, {Croton},
  {Porter}, {Primack}, {Behroozi}, {Ferguson}, {Koo}, {Guo}, {Safarzadeh},
  {Finlator}, {Castellano}, {White}, {Sommariva}, \& {Moody}}]{Lu2014}
{Lu}, Y., {Wechsler}, R.~H., {Somerville}, R.~S., {et~al.} 2014,
  \href{http://dx.doi.org/10.1088/0004-637X/795/2/123}{\JournalTitle{\apj},
  795, 123}

\bibitem[{{Ludlow} {et~al.}(2012){Ludlow}, {Navarro}, {Li}, {Angulo},
  {Boylan-Kolchin}, \& {Bett}}]{Ludlow2012}
{Ludlow}, A.~D., {Navarro}, J.~F., {Li}, M., {et~al.} 2012,
  \href{http://dx.doi.org/10.1111/j.1365-2966.2012.21892.x}{\JournalTitle{\mnras},
  427, 1322}

\bibitem[{{Mandelbaum} {et~al.}(2006){Mandelbaum}, {Seljak}, {Kauffmann},
  {Hirata}, \& {Brinkmann}}]{Mandelbaum2006}
{Mandelbaum}, R., {Seljak}, U., {Kauffmann}, G., {Hirata}, C.~M., \&
  {Brinkmann}, J. 2006,
  \href{http://dx.doi.org/10.1111/j.1365-2966.2006.10156.x}{\JournalTitle{\mnras},
  368, 715}

\bibitem[{{Mao} {et~al.}(2015){Mao}, {Williamson}, \& {Wechsler}}]{Mao2015}
{Mao}, Y.-Y., {Williamson}, M., \& {Wechsler}, R.~H. 2015,
  \href{http://dx.doi.org/10.1088/0004-637X/810/1/21}{\JournalTitle{\apj}, 810,
  21}

\bibitem[{{Mar{\'{\i}}n} {et~al.}(2008){Mar{\'{\i}}n}, {Wechsler}, {Frieman},
  \& {Nichol}}]{Marin2008}
{Mar{\'{\i}}n}, F.~A., {Wechsler}, R.~H., {Frieman}, J.~A., \& {Nichol}, R.~C.
  2008, \href{http://dx.doi.org/10.1086/523628}{\JournalTitle{\apj}, 672, 849}

\bibitem[{{Miyatake} {et~al.}(2015){Miyatake}, {More}, {Mandelbaum}, {Takada},
  {Spergel}, {Kneib}, {Schneider}, {Brinkmann}, \& {Brownstein}}]{Miyatake2015}
{Miyatake}, H., {More}, S., {Mandelbaum}, R., {et~al.} 2015,
  \href{http://dx.doi.org/10.1088/0004-637X/806/1/1}{\JournalTitle{\apj}, 806,
  1}

\bibitem[{{More} {et~al.}(2009){More}, {van den Bosch}, {Cacciato}, {Mo},
  {Yang}, \& {Li}}]{More2009}
{More}, S., {van den Bosch}, F.~C., {Cacciato}, M., {et~al.} 2009,
  \href{http://dx.doi.org/10.1111/j.1365-2966.2008.14095.x}{\JournalTitle{\mnras},
  392, 801}

\bibitem[{{More} {et~al.}(2016){More}, {Miyatake}, {Takada}, {Diemer},
  {Kravtsov}, {Dalal}, {More}, {Murata}, {Mandelbaum}, {Rozo}, {Rykoff},
  {Oguri}, \& {Spergel}}]{More2016}
{More}, S., {Miyatake}, H., {Takada}, M., {et~al.} 2016,
  \href{http://dx.doi.org/10.3847/0004-637X/825/1/39}{\JournalTitle{\apj}, 825,
  39}

\bibitem[{{Neistein} \& {Khochfar}(2012)}]{Neistein2012}
{Neistein}, E., \& {Khochfar}, S. 2012, \JournalTitle{ArXiv e-prints},
  \href{http://arxiv.org/abs/1209.0463}{{\sffamily arXiv:1209.0463
  [astro-ph.CO]}}

\bibitem[{{Neistein} {et~al.}(2011){Neistein}, {Li}, {Khochfar}, {Weinmann},
  {Shankar}, \& {Boylan-Kolchin}}]{Neistein2011}
{Neistein}, E., {Li}, C., {Khochfar}, S., {et~al.} 2011,
  \href{http://dx.doi.org/10.1111/j.1365-2966.2011.19145.x}{\JournalTitle{\mnras},
  416, 1486}

\bibitem[{{Norberg} {et~al.}(2009){Norberg}, {Baugh}, {Gazta{\~n}aga}, \&
  {Croton}}]{Norberg2009}
{Norberg}, P., {Baugh}, C.~M., {Gazta{\~n}aga}, E., \& {Croton}, D.~J. 2009,
  \href{http://dx.doi.org/10.1111/j.1365-2966.2009.14389.x}{\JournalTitle{\mnras},
  396, 19}

\bibitem[{{Padmanabhan} {et~al.}(2008){Padmanabhan}, {Schlegel}, {Finkbeiner},
  {Barentine}, {Blanton}, {Brewington}, {Gunn}, {Harvanek}, {Hogg},
  {Ivezi{\'c}}, {Johnston}, {Kent}, {Kleinman}, {Knapp}, {Krzesinski}, {Long},
  {Neilsen}, {Nitta}, {Loomis}, {Lupton}, {Roweis}, {Snedden}, {Strauss}, \&
  {Tucker}}]{Padmanabhan2008}
{Padmanabhan}, N., {Schlegel}, D.~J., {Finkbeiner}, D.~P., {et~al.} 2008,
  \href{http://dx.doi.org/10.1086/524677}{\JournalTitle{\apj}, 674, 1217}

\bibitem[{{Paranjape} {et~al.}(2015){Paranjape}, {Kova{\v c}}, {Hartley}, \&
  {Pahwa}}]{Paranjape2015}
{Paranjape}, A., {Kova{\v c}}, K., {Hartley}, W.~G., \& {Pahwa}, I. 2015,
  \href{http://dx.doi.org/10.1093/mnras/stv2137}{\JournalTitle{\mnras}, 454,
  3030}

\bibitem[{{Peacock} \& {Smith}(2000)}]{Peacock2000}
{Peacock}, J.~A., \& {Smith}, R.~E. 2000,
  \href{http://dx.doi.org/10.1046/j.1365-8711.2000.03779.x}{\JournalTitle{\mnras},
  318, 1144}

\bibitem[{{Reddick} {et~al.}(2013){Reddick}, {Wechsler}, {Tinker}, \&
  {Behroozi}}]{Reddick2013}
{Reddick}, R.~M., {Wechsler}, R.~H., {Tinker}, J.~L., \& {Behroozi}, P.~S.
  2013,
  \href{http://dx.doi.org/10.1088/0004-637X/771/1/30}{\JournalTitle{\apj}, 771,
  30}

\bibitem[{{Reid} {et~al.}(2014){Reid}, {Seo}, {Leauthaud}, {Tinker}, \&
  {White}}]{Reid2014}
{Reid}, B.~A., {Seo}, H.-J., {Leauthaud}, A., {Tinker}, J.~L., \& {White}, M.
  2014, \href{http://dx.doi.org/10.1093/mnras/stu1391}{\JournalTitle{\mnras},
  444, 476}

\bibitem[{{Rodr{\'{\i}}guez-Puebla} {et~al.}(2013){Rodr{\'{\i}}guez-Puebla},
  {Avila-Reese}, \& {Drory}}]{Rodriguez-Puebla2013}
{Rodr{\'{\i}}guez-Puebla}, A., {Avila-Reese}, V., \& {Drory}, N. 2013,
  \href{http://dx.doi.org/10.1088/0004-637X/767/1/92}{\JournalTitle{\apj}, 767,
  92}

\bibitem[{{Rodr{\'{\i}}guez-Puebla} {et~al.}(2012){Rodr{\'{\i}}guez-Puebla},
  {Drory}, \& {Avila-Reese}}]{Rodriguez-Puebla2012}
{Rodr{\'{\i}}guez-Puebla}, A., {Drory}, N., \& {Avila-Reese}, V. 2012,
  \href{http://dx.doi.org/10.1088/0004-637X/756/1/2}{\JournalTitle{\apj}, 756,
  2}

\bibitem[{{Rykoff} {et~al.}(2014){Rykoff}, {Rozo}, {Busha}, {Cunha},
  {Finoguenov}, {Evrard}, {Hao}, {Koester}, {Leauthaud}, {Nord}, {Pierre},
  {Reddick}, {Sadibekova}, {Sheldon}, \& {Wechsler}}]{Rykoff2014}
{Rykoff}, E.~S., {Rozo}, E., {Busha}, M.~T., {et~al.} 2014,
  \href{http://dx.doi.org/10.1088/0004-637X/785/2/104}{\JournalTitle{\apj},
  785, 104}

\bibitem[{{Saito} {et~al.}(2016){Saito}, {Leauthaud}, {Hearin}, {Bundy},
  {Zentner}, {Behroozi}, {Reid}, {Sinha}, {Coupon}, {Tinker}, {White}, \&
  {Schneider}}]{Saito2015}
{Saito}, S., {Leauthaud}, A., {Hearin}, A.~P., {et~al.} 2016,
  \href{http://dx.doi.org/10.1093/mnras/stw1080}{\JournalTitle{\mnras}, 460,
  1457}

\bibitem[{{Schaye} {et~al.}(2015){Schaye}, {Crain}, {Bower}, {Furlong},
  {Schaller}, {Theuns}, {Dalla Vecchia}, {Frenk}, {McCarthy}, {Helly},
  {Jenkins}, {Rosas-Guevara}, {White}, {Baes}, {Booth}, {Camps}, {Navarro},
  {Qu}, {Rahmati}, {Sawala}, {Thomas}, \& {Trayford}}]{2015MNRAS.446..521S}
{Schaye}, J., {Crain}, R.~A., {Bower}, R.~G., {et~al.} 2015,
  \href{http://dx.doi.org/10.1093/mnras/stu2058}{\JournalTitle{\mnras}, 446,
  521}

\bibitem[{{Scoccimarro} {et~al.}(2001){Scoccimarro}, {Sheth}, {Hui}, \&
  {Jain}}]{Scoccimarro2001}
{Scoccimarro}, R., {Sheth}, R.~K., {Hui}, L., \& {Jain}, B. 2001,
  \href{http://dx.doi.org/10.1086/318261}{\JournalTitle{\apj}, 546, 20}

\bibitem[{{Seljak}(2000)}]{Seljak2000}
{Seljak}, U. 2000,
  \href{http://dx.doi.org/10.1046/j.1365-8711.2000.03715.x}{\JournalTitle{\mnras},
  318, 203}

\bibitem[{{Skillman} {et~al.}(2014){Skillman}, {Warren}, {Turk}, {Wechsler},
  {Holz}, \& {Sutter}}]{Skillman2014}
{Skillman}, S.~W., {Warren}, M.~S., {Turk}, M.~J., {et~al.} 2014,
  \JournalTitle{ArXiv e-prints},
  \href{http://arxiv.org/abs/1407.2600}{{\sffamily arXiv:1407.2600}}

\bibitem[{{Somerville} \& {Dav{\'e}}(2015)}]{2015ARA&A..53...51S}
{Somerville}, R.~S., \& {Dav{\'e}}, R. 2015,
  \href{http://dx.doi.org/10.1146/annurev-astro-082812-140951}{\JournalTitle{\araa},
  53, 51}

\bibitem[{{Somerville} {et~al.}(2008){Somerville}, {Hopkins}, {Cox},
  {Robertson}, \& {Hernquist}}]{Somerville2008}
{Somerville}, R.~S., {Hopkins}, P.~F., {Cox}, T.~J., {Robertson}, B.~E., \&
  {Hernquist}, L. 2008,
  \href{http://dx.doi.org/10.1111/j.1365-2966.2008.13805.x}{\JournalTitle{\mnras},
  391, 481}

\bibitem[{{Springel}(2005)}]{Springel2005}
{Springel}, V. 2005,
  \href{http://dx.doi.org/10.1111/j.1365-2966.2005.09655.x}{\JournalTitle{\mnras},
  364, 1105}

\bibitem[{{Sunayama} {et~al.}(2016){Sunayama}, {Hearin}, {Padmanabhan}, \&
  {Leauthaud}}]{Sunayama2015}
{Sunayama}, T., {Hearin}, A.~P., {Padmanabhan}, N., \& {Leauthaud}, A. 2016,
  \href{http://dx.doi.org/10.1093/mnras/stw332}{\JournalTitle{\mnras}, 458,
  1510}

\bibitem[{{Tasitsiomi} {et~al.}(2004){Tasitsiomi}, {Kravtsov}, {Wechsler}, \&
  {Primack}}]{Tasitsiomi2004}
{Tasitsiomi}, A., {Kravtsov}, A.~V., {Wechsler}, R.~H., \& {Primack}, J.~R.
  2004, \href{http://dx.doi.org/10.1086/423784}{\JournalTitle{\apj}, 614, 533}

\bibitem[{{Tinker} {et~al.}(2011){Tinker}, {Wetzel}, \& {Conroy}}]{Tinker2011}
{Tinker}, J., {Wetzel}, A., \& {Conroy}, C. 2011, \JournalTitle{ArXiv
  e-prints}, \href{http://arxiv.org/abs/1107.5046}{{\sffamily arXiv:1107.5046
  [astro-ph.CO]}}

\bibitem[{{Tinker} \& {Conroy}(2009)}]{Tinker2009}
{Tinker}, J.~L., \& {Conroy}, C. 2009,
  \href{http://dx.doi.org/10.1088/0004-637X/691/1/633}{\JournalTitle{\apj},
  691, 633}

\bibitem[{{Tinker} {et~al.}(2008){Tinker}, {Conroy}, {Norberg}, {Patiri},
  {Weinberg}, \& {Warren}}]{Tinker2008}
{Tinker}, J.~L., {Conroy}, C., {Norberg}, P., {et~al.} 2008,
  \href{http://dx.doi.org/10.1086/589983}{\JournalTitle{\apj}, 686, 53}

\bibitem[{{Tinker} {et~al.}(2012){Tinker}, {George}, {Leauthaud}, {Bundy},
  {Finoguenov}, {Massey}, {Rhodes}, \& {Wechsler}}]{Tinker2012}
{Tinker}, J.~L., {George}, M.~R., {Leauthaud}, A., {et~al.} 2012,
  \href{http://dx.doi.org/10.1088/2041-8205/755/1/L5}{\JournalTitle{\apjl},
  755, L5}

\bibitem[{{Trujillo-Gomez} {et~al.}(2011){Trujillo-Gomez}, {Klypin}, {Primack},
  \& {Romanowsky}}]{Trujillo-Gomez2011}
{Trujillo-Gomez}, S., {Klypin}, A., {Primack}, J., \& {Romanowsky}, A.~J. 2011,
  \href{http://dx.doi.org/10.1088/0004-637X/742/1/16}{\JournalTitle{\apj}, 742,
  16}

\bibitem[{{Vakili}(2016)}]{1610:01991}
{Vakili}, M. 2016, \JournalTitle{ArXiv e-prints},
  \href{http://arxiv.org/abs/1610.01991}{{\sffamily arXiv:1610.01991}}

\bibitem[{{Vale} \& {Ostriker}(2004)}]{Vale2004}
{Vale}, A., \& {Ostriker}, J.~P. 2004,
  \href{http://dx.doi.org/10.1111/j.1365-2966.2004.08059.x}{\JournalTitle{\mnras},
  353, 189}

\bibitem[{{Vale} \& {Ostriker}(2006)}]{Vale2006}
---. 2006,
  \href{http://dx.doi.org/10.1111/j.1365-2966.2006.10605.x}{\JournalTitle{\mnras},
  371, 1173}

\bibitem[{{Vogelsberger} {et~al.}(2014){Vogelsberger}, {Genel}, {Springel},
  {Torrey}, {Sijacki}, {Xu}, {Snyder}, {Nelson}, \&
  {Hernquist}}]{Vogelsberger2014}
{Vogelsberger}, M., {Genel}, S., {Springel}, V., {et~al.} 2014,
  \href{http://dx.doi.org/10.1093/mnras/stu1536}{\JournalTitle{\mnras}, 444,
  1518}

\bibitem[{Warren(2013)}]{Warren2013}
Warren, M.~S. 2013, \href{http://dx.doi.org/10.1145/2503210.2503220}{in
  Proceedings of the International Conference on High Performance Computing,
  Networking, Storage and Analysis, SC '13} (New York, NY, USA: ACM), 72:1

\bibitem[{{Wechsler} {et~al.}(2001){Wechsler}, {Somerville}, {Bullock},
  {Kolatt}, {Primack}, {Blumenthal}, \& {Dekel}}]{Wechsler2001}
{Wechsler}, R.~H., {Somerville}, R.~S., {Bullock}, J.~S., {et~al.} 2001,
  \href{http://dx.doi.org/10.1086/321373}{\JournalTitle{\apj}, 554, 85}

\bibitem[{{Wechsler} {et~al.}(2006){Wechsler}, {Zentner}, {Bullock},
  {Kravtsov}, \& {Allgood}}]{Wechsler2006}
{Wechsler}, R.~H., {Zentner}, A.~R., {Bullock}, J.~S., {Kravtsov}, A.~V., \&
  {Allgood}, B. 2006,
  \href{http://dx.doi.org/10.1086/507120}{\JournalTitle{\apj}, 652, 71}

\bibitem[{{Yang} {et~al.}(2006){Yang}, {Mo}, \& {van den Bosch}}]{Yang2006}
{Yang}, X., {Mo}, H.~J., \& {van den Bosch}, F.~C. 2006,
  \href{http://dx.doi.org/10.1086/501069}{\JournalTitle{\apjl}, 638, L55}

\bibitem[{{Zehavi} {et~al.}(2011){Zehavi}, {Zheng}, {Weinberg}, {Blanton},
  {Bahcall}, {Berlind}, {Brinkmann}, {Frieman}, {Gunn}, {Lupton}, {Nichol},
  {Percival}, {Schneider}, {Skibba}, {Strauss}, {Tegmark}, \&
  {York}}]{Zehavi2011}
{Zehavi}, I., {Zheng}, Z., {Weinberg}, D.~H., {et~al.} 2011,
  \href{http://dx.doi.org/10.1088/0004-637X/736/1/59}{\JournalTitle{\apj}, 736,
  59}

\bibitem[{{Zentner} {et~al.}(2005){Zentner}, {Berlind}, {Bullock}, {Kravtsov},
  \& {Wechsler}}]{Zentner2005}
{Zentner}, A.~R., {Berlind}, A.~A., {Bullock}, J.~S., {Kravtsov}, A.~V., \&
  {Wechsler}, R.~H. 2005,
  \href{http://dx.doi.org/10.1086/428898}{\JournalTitle{\apj}, 624, 505}

\bibitem[{{Zentner} {et~al.}(2016){Zentner}, {Hearin}, {van den Bosch},
  {Lange}, \& {Villarreal}}]{1606.07817}
{Zentner}, A.~R., {Hearin}, A., {van den Bosch}, F.~C., {Lange}, J.~U., \&
  {Villarreal}, A. 2016, \JournalTitle{ArXiv e-prints},
  \href{http://arxiv.org/abs/1606.07817}{{\sffamily arXiv:1606.07817}}

\bibitem[{{Zentner} {et~al.}(2014){Zentner}, {Hearin}, \& {van den
  Bosch}}]{Zentner2014}
{Zentner}, A.~R., {Hearin}, A.~P., \& {van den Bosch}, F.~C. 2014,
  \href{http://dx.doi.org/10.1093/mnras/stu1383}{\JournalTitle{\mnras}, 443,
  3044}

\end{thebibliography}

\end{document}